\documentclass[aps,prr,twocolumn,superscriptaddress]{revtex4}
\usepackage{graphicx}
\usepackage{dcolumn}
\usepackage{bm}
\usepackage{physics}
\usepackage{amsmath}
\usepackage{amssymb}
\usepackage{array}
\usepackage{color}
\usepackage{xcolor}
\newcolumntype{P}[1]{>{\centering\arraybackslash}p{#1}}
\usepackage[colorlinks=true, breaklinks=true, linkcolor=blue, citecolor=blue, urlcolor=blue]{hyperref}
\newcommand{\CsCoBr}{Cs$_2$CoBr$_4$\xspace}
\newcommand{\CsCuCl}{Cs$_2$CuCl$_4$\xspace}
\newcommand{\mb}[1]{\mathbf{#1}}

\newcommand{\geneva}{Department of Quantum Matter Physics, University of Geneva, Quai Ernest-Ansermet 24, 1211 Geneva, Switzerland}
\newcommand{\ethz}{Laboratory for Solid State Physics, ETH Z\"{u}rich, 8093 Z\"{u}rich, Switzerland}
\newcommand{\pks}{Max Planck Institute for the Physics of Complex Systems, N\"othnitzer Str.~38, 01187 Dresden, Germany}

\usepackage{ifthen}
\usepackage{booktabs}
\usepackage{multirow}

\usepackage{amsmath,amssymb,bm,graphicx,color,gensymb,bbold,appendix,xspace}

\begin{document}

\title{Zeeman Ladders in Frustrated XYZ Spin Chains}

\date{\today}

\begin{abstract}
We investigate the nature of the excitations captured by the dynamical response of XYZ triangular spin-1/2 ladders. We complement experimental inelastic neutron scattering results on the compound Cs\textsubscript{2}CoBr\textsubscript{4} with numerically exact simulations based on time-dependent matrix product state methods.
Our results show that bound states of spinon excitations can arise in XYZ beyond the requirement of strong Ising anisotropies.
We analyze the role of the frustrated triangular couplings on the excitations giving rise to the spin dynamical structure factor and show how the features of the bound states manifest themselves in the different polarization channels.
\end{abstract}
\author{Catalin-Mihai Halati}
\affiliation{\pks}
\affiliation{\geneva}
\author{Viola Romerio}
\affiliation{\ethz}
\author{Paul Steffens}
\affiliation{Institut Laue-Langevin, 71 Avenue des Martyrs, CS 20156, 38042 Grenoble Cedex 9, France}
\author{J.~Ross Stewart}
\affiliation{ISIS Neutron and Muon Source, Rutherford Appleton
Laboratory, Didcot, OX11 0QX, United Kingdom}
\author{Andrey Zheludev}
\affiliation{\ethz}
\author{Thierry Giamarchi}
\affiliation{\geneva}
\maketitle

\section{Introduction}

Quantum magnetic insulators show an extremely rich physics \cite{Auerbach1994, Mattis2006, Starykh2015, VasilievMarkina2018, SavaryBalents2017, ZhouNg2017}, ranging from ordered states to spin liquids, stemming from the interplay of the geometry of the material and the nature of the magnetic exchange couplings. Understanding the underlying mechanisms and the resulting quantum phenomena is one of the extremely challenging problems of the field.
In recent years, the synergy between experimental and theoretical advancements has led to great success in uncovering the unusual physics occurring in low-dimensional quantum spin systems, as for example, various aspects of Tomonaga-Luttinger liquid theory \cite{KlanjsekGiamarchi2008, BouillotGiamarchi2011, SchmidigerZheludev2013, FaurePetit2019, CuiYu2022}, topological phase transitions \cite{FaureGrenier2018, CuiYu2019}, high energy bound states \cite{ColdeaKiefer2010, MorrisArmitage2014, WangLoidl2018, BeraLake2020, HalatiBernier2023, WangKollath2024}, and signatures of supersymmetry \cite{WehingerRuegg2025}.

Due to their unique topology, one-dimensional (1D) spin systems
often host exotic quasi-particles with fractional quantum
numbers, known as spinons. When spinons are unbound, as in spin $S=1/2$ Heisenberg chains, they give rise to structured multi-particle
excitation continua \cite{descloiseaux_spectrum_spinchain, StoneTurnbull2003, ZaliznyakTakagi2004, ThielemannMesot2009b, MourigalRonnow2013, LakeFrost2013}. 
Even more peculiar physics emerges from spinon confinement. 
The latter can be topological in nature, e.g.~in $S=1$ Haldane spin chains \cite{GreiterRachel2007, RachelGreiter2009, VanderstraetenMila2020}, but can also be caused by Weiss fields, stemming from higher dimensional couplings in quasi-1D magnetic materials, or externally applied fields. 
The result is a complicated hierarchy of stable and mobile bound states known as ``Zeeman ladders''.
The concept originates in the Ising XXZ spin chains, where spinons can be semiclassically pictured as ``kinks'' separating magnetic domains \cite{McCoyWu1978, Shiba1980, Rutkevich2008}. 
A field which favors one type of domain over the other determines an effective attractive interaction between the spinons, leading to confinement. 
This scenario has found realization in several magnetic materials, where neutron spectroscopy and other experimental probes provided spectacular direct observations of the spinon bound states \cite{ColdeaKiefer2010, GrenierLejay2015, WangLoidl2016,FaureGrenier2018, BeraQuinteroCastro2017, MenaRuegg2020, MatanSato2022, GhioldiBatista2022, FacherisZheludev2023}. 
The natural question that follows is how robust such spectral features are and to what extent do they rely on angular momentum conservation? 
While most of such studies analyzed the confinement of spinons in unfrustrated geometries, it remained an open question how the presence of frustration influences the above-mentioned physics.

First hints towards answering these questions were obtained in measurements on the antiferromagnetic insulator \CsCoBr \cite{PovarovZheludev2020, FacherisZheludev2022, FacherisZheludev2023, FacherisZheludev2024}. 
Its low-energy physics can be captured by effective spin-$1/2$ degrees of freedom with general XYZ interactions. The distorted triangular lattice geometry and the particular arrangement of anisotropy axes produce a great deal of frustration in this compound. 
Inelastic neutron scattering (INS) experiments revealed a highly structured Zeeman ladder of magnetic bound states \cite{FacherisZheludev2024} that differs significantly from those in Ising spin chains \cite{FaureGrenier2018,ColdeaKiefer2010, GrenierLejay2015, WangLoidl2016, BeraQuinteroCastro2017, MenaRuegg2020}. 

In this work, we tackle the complex behavior of spinon confinement in frustrated chains 
with general anisotropies, by combining theoretical methods for low-dimensional quantum systems with additional INS experiments. 
We capture very well the dynamical response of \CsCoBr with a model of frustrated XYZ triangular spin ladders. Our results show that the frustrated geometry has an important impact on how the spinon bound states manifest in the dynamical spin response in both the simulations and the experiment. 
We further analyze how the Zeeman ladders in \CsCoBr stem from the contributions of the different spin polarizations to the dynamical spin structure factor.

\section{Triangular spin ladder model}

\begin{figure}[!hbtp]
\centering
\includegraphics[width=0.45\textwidth]{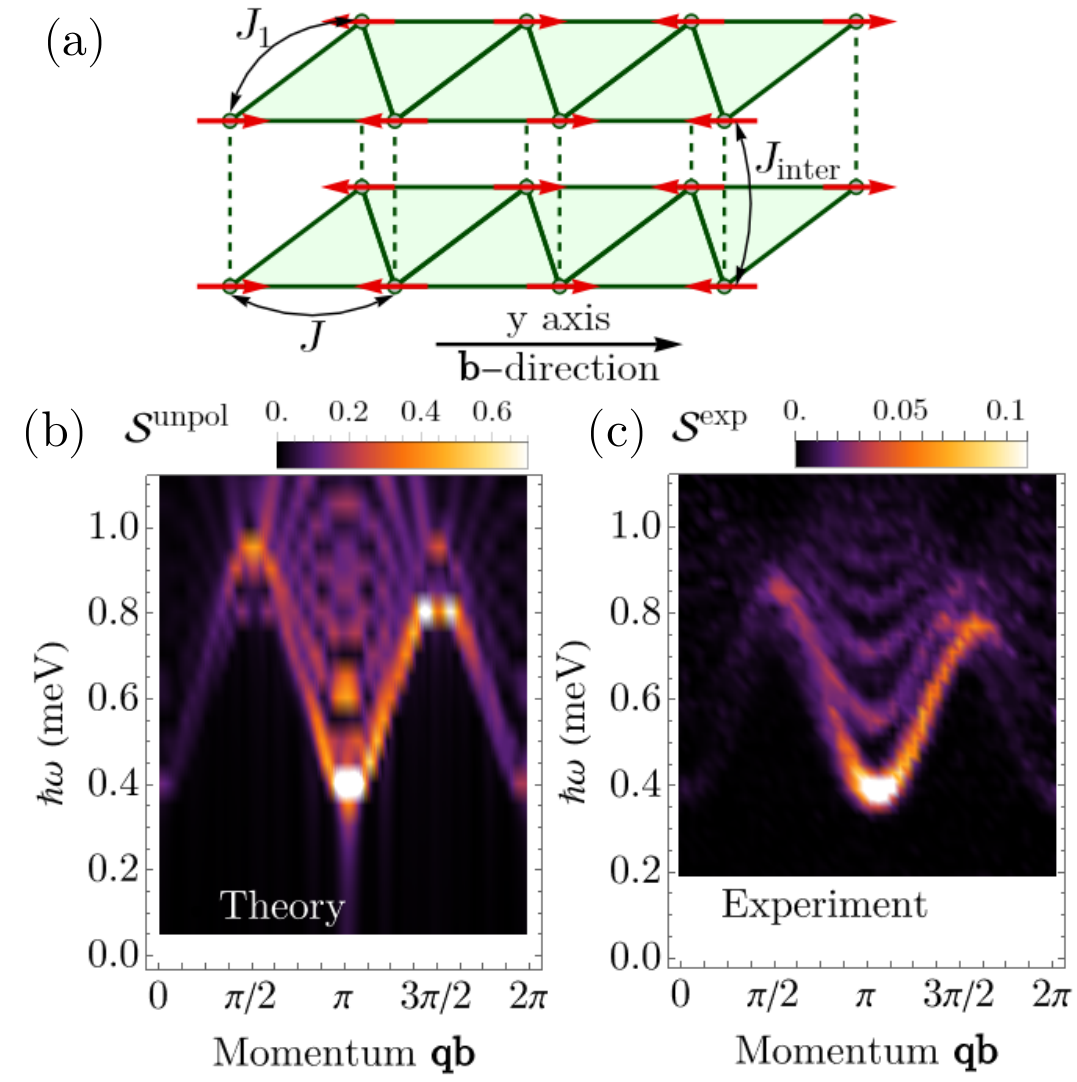}
\caption{\label{fig:sketch_SSF_unpol}
(a) Sketch of two coupled triangular ladders, with $J$ and $J_1$ interactions along the legs and rungs, respectively. In Cs\textsubscript{2}CoBr\textsubscript{4} the legs run along the $y$ axis (crystallographic $\bm{b}$ direction) for all ladders. The planes of adjacent ladders are almost orthogonal and form almost $45^\circ$ angles with respect to the $\bm{a}$ and $\bm{c}$ axes (illustrated in Fig.~1 of Ref.~\cite{PovarovZheludev2020}). $J_\text{inter}$ represents the coupling of the two ladders in a square geometry, which is a simplification of the real compound \cite{FacherisZheludev2024}.  The red arrows sketch the local expectation values of $\langle S^y \rangle$ in a N\'eel pattern.
(b) Numerical results for the dynamical spin structure factor $\mathcal{S}^\text{unpol}$ (see \ref{sec:app_SSF} for the definition). The parameters used are: $L=128$ sites, $J_1/J=0.45$, $\delta=0.6$, $J_\text{inter}/J=0.125$, $\Delta_z=0.25$,  $J=0.5~\text{meV}$, with the momentum transfer $\mathbf{q}=(0,k,0.5)$.
(c) Inelastic unpolarized neutron scattering experimental results for the compound Cs\textsubscript{2}CoBr\textsubscript{4}, taken from Ref.~\cite{FacherisZheludev2023, dataINS} for $\mathbf{q}=(0,k,0.5)$.
}
\end{figure}

Originally, by analogy with the well-known isostructural compound \CsCuCl \cite{ColdeaTylczynski2001, TokiwaSteglich2006, StarykhBalents2010, SmirnovShapiro2012}, \CsCoBr was believed to realize a distorted triangular spin lattice \cite{PovarovZheludev2020, FacherisZheludev2022}. 
However, subsequent high-field INS experiments \cite{FacherisZheludev2024} revealed a very different picture. 
The material is modeled by magnetic $S=3/2$ Co$^{2+}$ ions coupled through an hierarchy of at least five Heisenberg exchange interactions and subject to general XYZ single-ion anisotropy. 
The complexity of the frustrated three-dimensional model makes direct numerical studies unfeasible, thus, we introduce a minimal model which captures the main experimental features.
We first note that the two strongest interactions form triangular ladders, which can also be viewed as spin chains with competing nearest- and next-nearest neighbor antiferromagnetic couplings. 
The ladders run along the $\mathbf{b}$ crystallographic axis, with a spacing $b$ between consecutive spins in each leg. 
We base our simplified model on this 1D structure, sketched in Fig.~\ref{fig:sketch_SSF_unpol}(a). 
The coupling between different ladders is considered at a mean-field level, specifically, all interactions are replace by a single static Weiss field stemming from the magnetic long range order appearing in \CsCoBr below $T_\text{N}=1.3~\text{K}$ \cite{PovarovZheludev2020, FacherisZheludev2022}. 
Additionally, the mean-field decoupling makes the complicated arrangement of local anisotropy frames of
different ions in \CsCoBr irrelevant, as all ions within one triangular ladder are identical by crystal symmetry.
In the complete spin-$3/2$ model, the single-ion anisotropy is rather large and predominantly of easy-plane type \cite{FacherisZheludev2024}, allowing us to significantly reduce the Hilbert space by employing a Schrieffer-Wolff projection to describe the low-energy physics in terms of effective $S=1/2$ pseudo-spin degrees of freedom \cite{BreunigLorenz2013, BreunigLorenz2015} (see \ref{app_model}).
In contrast to previous works, we replace the original single-ion anisotropy by anisotropic exchange. The resulting Hamiltonian for a single triangular ladder is
\begin{align}
\label{eq:Hamiltonian}
H_\text{XYZ} &=H_\text{leg}+H_\text{rung} \\
H_\text{leg}&=J\sum_{j,m} \big[S_{m,j}^x S^x_{m,j+1}+(1+\delta) S_{m,j}^y S^y_{m,j+1} \nonumber \\
&\quad+\Delta_z S_{m,j}^z S^z_{m,j+1}\big] \nonumber \\
H_\text{rung}&=J_1\sum_{j} \big[S_{1,j}^x S^x_{2,j} +(1+\delta) S_{1,j}^y S^y_{2,j}+\Delta_z S_{1,j}^z S^z_{2,j} \nonumber  \\
&+S_{1,j+1}^x S^x_{2,j}+(1+\delta) S_{1,j+1}^y S^y_{2,j}+ \Delta_z S_{1,j+1}^z S^z_{2,j}\big]. \nonumber
\end{align}
Here $H_\text{leg}$ and $H_\text{rung}$ describe interactions in the legs and zig-zag rungs, respectively, with $m\in\{1,2\}$ labeling the two legs. The coordinate system is as used in Ref.~\cite{PovarovZheludev2020}, with the ladder legs along the $y$ direction (crystallographic $\mathbf{b}$-axis).
The $x$ and $z$ axes are in the $(\mathbf{a},\mathbf{c})$ crystallographic plane, roughly at 45$^\circ$ relative the crystallographic $\mathbf{a}$ axis.
The couplings $J, J_1>0$ correspond to antiferromagnetic
interactions (labeled as $J_4$ and $J_1$ in Ref.~\cite{FacherisZheludev2024}).  The parameters $0<\Delta_z<1$ and $\delta > 0$ determine the strength of the easy-plane and weak in-plane anisotropies, correspondingly.

Inter-ladder interactions are essential for long-range magnetic ordering in \CsCoBr and for the formation of bound states. 
We treat these interactions in a mean-field approach, 
$H_\text{MF}=-J_\text{inter}\sum_{m,j} \langle S_{m,j}^y \rangle S^y_{m,j}$ (see \ref{sec:app_gs}), where $J_\text{inter}$ is the non-frustrated part of inter-ladder coupling.
The anisotropy geometry leads to ordered moments along the $y$-direction, both in \CsCoBr and the ladder model, Eq.~(\ref{eq:Hamiltonian}), with a N\'eel-type structure illustrated in Fig.~\ref{fig:sketch_SSF_unpol}(a).
The interladder coupling becomes an effective staggered magnetic field along each chain, leading to the Hamiltonian
\begin{align}
\label{eq:Hamiltonian_MF}
H=H_\text{XYZ}+H_\text{MF},~H_\text{MF}=-h\sum_{m,j} (-1)^{j+m} S^y_{m,j},
\end{align}
where the Weiss field $h\equiv J_\text{inter}\sum_{m,j} \langle S_{m,j}^y\rangle/L$ is determined self-consistently in the ground state of the model (see \ref{sec:app_gs}).

\section{Results}

We study the impact of the frustrated geometry and the anisotropy on the confinement dynamics of the spinons as reflected in the dynamical response of the spin model and measurements on the compound \CsCoBr.
We determine the excitation spectrum by numerically computing the dynamical spin structure factor, $\mathcal{S}^{\alpha\alpha}(\mathbf{q},\omega)$.
The structure factor is defined as the spatial and temporal Fourier transform of the spin-spin correlation function $\langle S^{\alpha}_{j}(t)S^{\alpha}_{l}(0)\rangle$ and is numerically evaluated using the time-dependent matrix-product-state (tMPS) algorithm \cite{WhiteFeiguin2004, DaleyVidal2004, Schollwoeck2011} (see \ref{app_methods}).
The detailed procedure and how to related the experimentally measured INS signal to the numerically determined correlations are described in \ref{sec:app_SSF}.
The ground state was computed numerically using the density matrix renormalization group (DMRG) algorithm \cite{White1992, Schollwoeck2011}.
We represent our results as a function of the projection of the momentum onto the $\mathbf{b}$-axis, $\mathbf{qb}$, the component of the three-dimensional vector that is experimentally varied.

Our simplified spin-$1/2$ triangular ladder model, Eq.~(\ref{eq:Hamiltonian})-(\ref{eq:Hamiltonian_MF}), seems to capture the behavior observed in  \CsCoBr remarkably well. 
We determine the scattering intensity for an unpolarized neutron beam \cite{Lovesey1984, Squires2012}, by computing the dynamical structure factors for the actual three-dimensional arrangement of magnetic ions in the crystal, while taking into account the polarization factors for unpolarized neutrons \cite{Squires2012} and the magnetic form factor of Co$^{2+}$ \cite{Brown2001} (see \ref{sec:app_SSF}). 
The numerical result is plotted in Fig.~\ref{fig:sketch_SSF_unpol}(b), next to the experimental data from Ref.~\cite{FacherisZheludev2024} in Fig.~\ref{fig:sketch_SSF_unpol}(c).
The values of the parameters, $J_1/J=0.45$, $\delta=0.6$, $J_\text{inter}/J=0.125$, $\Delta_z=0.25$, $J=0.5~\text{meV}$,   have been optimized such that the gap and the bandwidth are similar with the experimental ones. They compare reasonably with the ones employed in Ref.~\cite{FacherisZheludev2024} where the INS measurements were compared with spin wave theory results of a related spin $3/2$ model.
The main contributions to the experimental spectrum: the gapped lower band, the Zeeman ladder structure of the spectrum, the asymmetry of the dispersion and the overall intensity distribution are reproduced by the numerical results at a semi-quantitative level.
In the following, we show that while the gapped ladder of excitations can be understood by considering decoupled single chains, the momentum asymmetry stems from the triangular couplings.

\begin{figure}[!hbtp]
\centering
\includegraphics[width=0.48\textwidth]{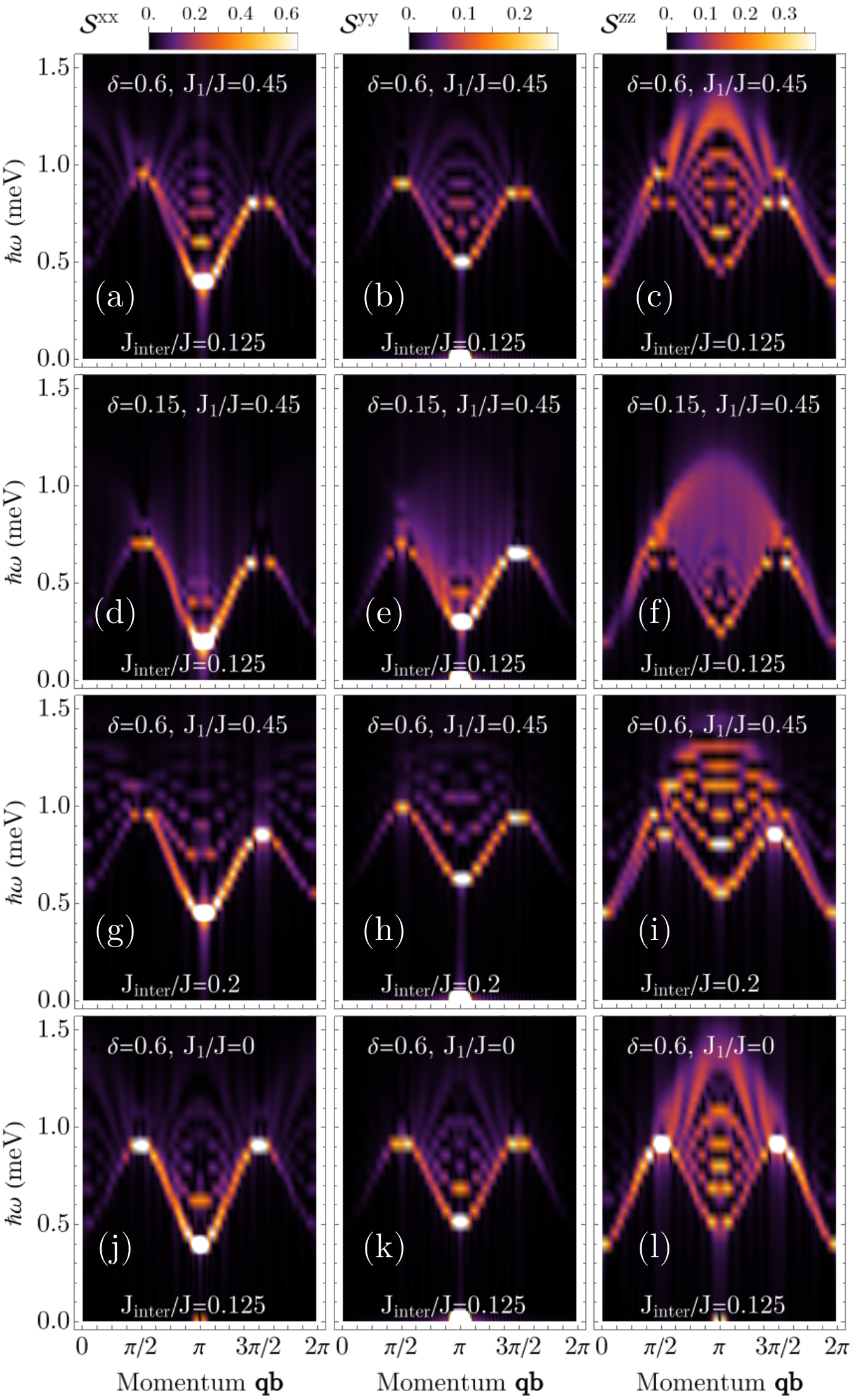}
\caption{\label{fig:SSF_numerics}
The dynamical spin structure factor as a function of momentum $q$ and frequency $\omega$ for the three spin directions polarization, $\mathcal{S}^{xx}(q,\omega)$,  $\mathcal{S}^{yy}(q,\omega)$ and  $\mathcal{S}^{zz}(q,\omega)$, using the Hamiltonian $H$, Eq.~(\ref{eq:Hamiltonian_MF}), for different values of the parameters of the model, $\delta$, $J_1/J$ and $J_\text{inter}/J$.
The other parameters employed are $L=128$ sites, $\Delta_z=0.25$,  $J=0.5~\text{meV}$, with momentum transfer $\mathbf{q}=(0,1,0)$.
}
\end{figure}

The ground state of the symmetric XXZ limit, $\delta=0$, of the Hamiltonian Eq.~(\ref{eq:Hamiltonian}), is gapless with antiferomagnetic correlations along the chains of the ladder, for $J_1\geq 0$.
However, a finite value of $\delta$ breaks this symmetry and induced a gap in the spectrum
[Fig.~\ref{fig:sketch_SSF_unpol}(b)-Fig.~\ref{fig:sketch_SSF_unpol}(c)].
This is also visible in the dynamical structure factor simulations shown in Fig.~\ref{fig:SSF_numerics}, where we separately show the longitudinal part $\mathcal{S}^{yy}(\mathbf{q},\omega)$, which represents spin fluctuations along the ordered moment, and the transverse components $\mathcal{S}^{xx}(\mathbf{q},\omega)$ and $\mathcal{S}^{zz}(\mathbf{q},\omega)$.
The gap has a similar value both in the presence and the absence of the triangular couplings [{\em cf.} Figs.~\ref{fig:SSF_numerics}(j)-(l) and Figs.~\ref{fig:SSF_numerics}(a)-(c)].
We observe the reduction of the excitation gap as we reduce the value the anisotropy to $\delta=0.15$ [Figs.~\ref{fig:SSF_numerics}(d)-(f)].

The $\delta$ anisotropy plays the role of an Ising-type term and determines the N\'eel order with a finite staggered magnetization along the $y$-direction (see \ref{sec:app_gs}), consistent with the diffraction studies in Ref.~\cite{FacherisZheludev2022}.
In a naive semiclassical analysis one would expect that the N\'eel order would decouple the two chains, as each site would be coupled to both a spin up and a spin down from the other leg.  
However, guided by field theoretical arguments we identified in the ground-state DMRG simulations four-spin correlations on the rungs whose finite value cannot be captured at the semiclassical level, as detailed in \ref{sec:app_gs}.
Furthermore, we argue that the fluctuations beyond the semiclassical limit are also relevant for the dynamical response of the spinon bound states.

\begin{figure}[!hbtp]
\centering
\includegraphics[width=0.48\textwidth]{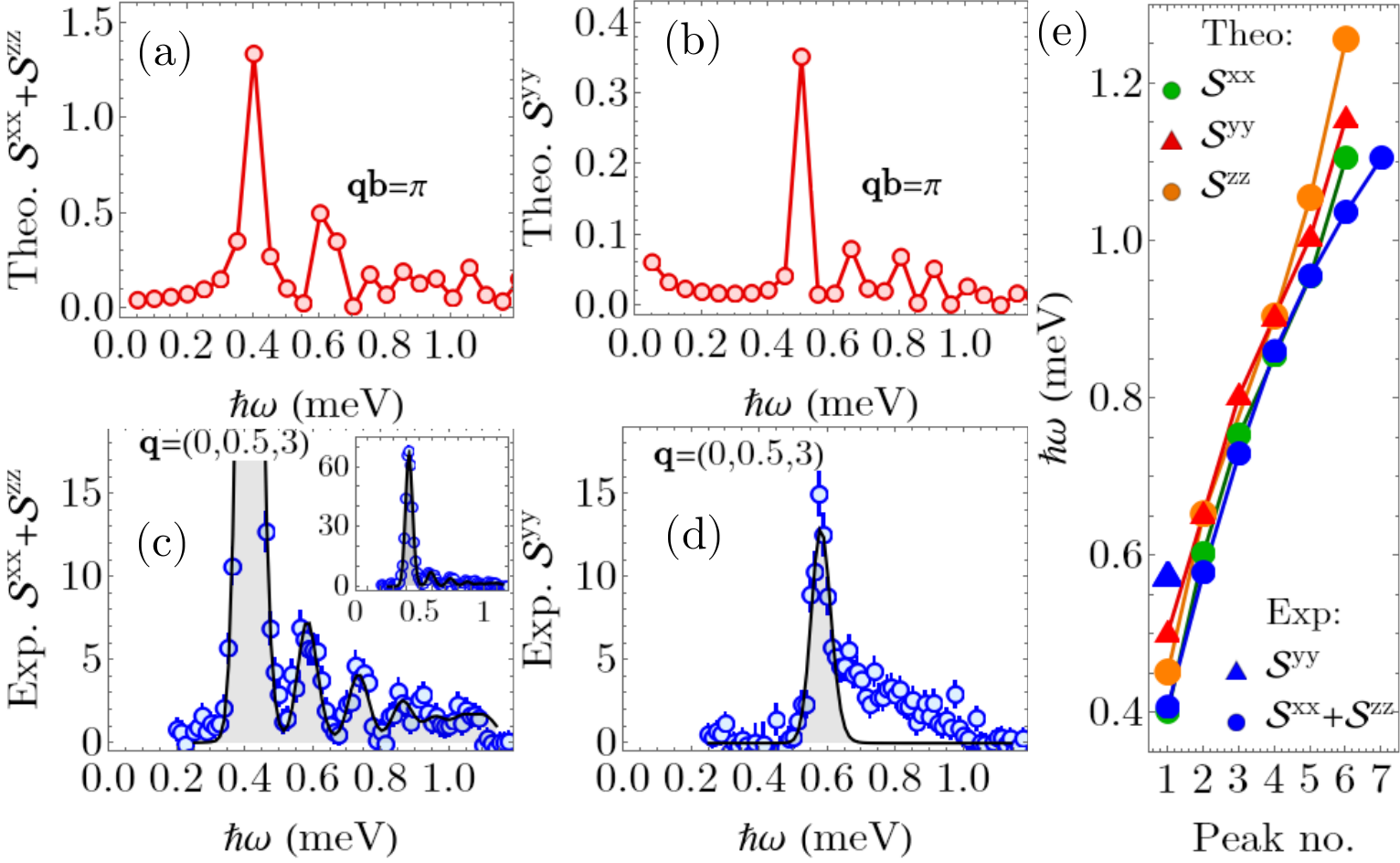}
\caption{\label{fig:SSF_cuts}
(a)-(b) Frequency dependence of the numerical dynamical spin structure factors (a) $\mathcal{S}^{yy}$ and (b) $(\mathcal{S}^{xx}+\mathcal{S}^{zz})$ at $\mathbf{qb}=\pi$, for $\delta=0.6$, $\Delta_z=0.25$, $J_\text{inter}/J=0.125$, $J=0.5~\text{meV}$, $L=128$ sites.
The momentum transfer considered is $\mathbf{q}=(0,0.5,3)$, as in the experiment.
(c)-(d) Experimental intensity (arbitrary units) for inelastic polarized neutron scattering measurements for the compound Cs\textsubscript{2}CoBr\textsubscript{4}, distinguishing between (c) the excitations in the direction of the chains, $y$ axis, and (d) the excitations perpendicular to the chain axis, $x$-$z$ plane.
(e) The extracted energies of the bound states for the inelastic polarized neutron scattering and the three theoretical polarizations.
}
\end{figure}

In all polarization channels the most prominent feature is a sharp mode on the lower bound of the spectrum that is followed by a series of dispersive weaker modes -- the Zeeman ladder, Fig.~\ref{fig:SSF_numerics}(a)-(c).
While for transverse spin components one could naively interpret the lowest-energy mode as a ``spin wave'', the description is of course inadequate for the longitudinal mode or any of the higher-energy excitations.
A more appropriate terminology is two-spinon bound states \cite{McCoyWu1978, Shiba1980, Rutkevich2008}. 
If the ladders were totally decoupled from their neighbors, the cost of creating two spinons would be $\propto J(1+\delta)$, which would be able to freely propagate. 
The dominant spectral feature would be a continuum of excitations similar to the famous ``butterfly spectrum'' of the XXZ Ising chain \cite{descloiseaux_spectrum_spinchain, MuellerBonner1981, KarbachMutter1997, Giamarchibook}, in particular without any peaks signaling the presence of bound states.
The presence of the other ladders and the mean-field Weiss field term, Eq.~(\ref{eq:Hamiltonian_MF}), effectively produces a linear potential between the spinons and leads to their confinement \cite{McCoyWu1978, Shiba1980, Rutkevich2008}.
Consequently, the spacing between bounds states and their spectral weight increases for larger values of $J_\text{inter}$ [{\em cf.}~Figs.~\ref{fig:SSF_numerics}(a)-(c) and Figs.~\ref{fig:SSF_numerics}(g)-(i)], consistent with the presence of a stronger confining potential.
This demonstrates the importance of the interladder coupling, both in the numerical and experimental results, as the triangular coupling between the two legs of a ladder is not sufficient to induce the spinon confinement by itself.
Furthermore, in Figs.~\ref{fig:SSF_numerics}(a)-(c) for $\delta=0.6$, we can clearly identify the features coming from multiple bound states, however, if we lower the anisotropy to $\delta=0.15$, as in Figs.~\ref{fig:SSF_numerics}(d)-(f), the presence of bound states becomes much more difficult to ascertain.
This effect comes from the fact that the energy separation between the supspaces with multiple spinons on a leg of the ladder is controlled by $\delta$, thus, a larger value entails a suppression of the processes connecting states with different number of spinons, facilitating the confinement. 
Thus, $\delta$ corresponds to the large Ising anisotropy of the previous observations of spinon bound states \cite{ColdeaKiefer2010, GrenierLejay2015, WangLoidl2016, BeraQuinteroCastro2017, MenaRuegg2020}.

In order to understand the influence of the frustrated triangular coupling on the spinon confinement it is useful to think about the semiclassical limit. In this limit, the spinon bound state consists in two domain walls at a certain distance on the same leg. 
On the other leg only the two spins which are at the same position as the domain walls are affected by the presence of the bound state as an additional potential of $\pm J_1 (1+\delta)$ is induced (see \ref{sec:semiclassical}). 
This potential has opposite signs for the first and the second domain wall, but it is not sufficient to induce spinons on the second leg, which would cost $J (1+\delta)$.
Thus, within this approximation there is no difference between the spinon bound state on the triangular ladder and on a chain.
However, if we compare the structure factor for the case of decoupled chains, [Fig.~\ref{fig:SSF_numerics}(j)-(l)], and the triangular ladder, [Fig.~\ref{fig:SSF_numerics}(a)-(c)], we observe that while the energies of the bound states at  $\mathbf{qb}=\pi$ are similar, the dispersion of the bands away from  $\mathbf{qb}=\pi$ is very different. The decoupled chain exhibits a symmetry of the correlations around $\mathbf{qb}=\pi$, which is not present in the presence of the triangular couplings.
Importantly, the experimental results shown in Fig.~\ref{fig:sketch_SSF_unpol}(c) exhibit this asymmetry, thus, the frustrated triangular is needed in order to capture the excitation spectrum at all momenta.
We attribute this feature of the dynamical response to quantum fluctuations beyond the semiclassical picture, in particular, stemming from processes taking place on the rungs of the ladder.
Due to the general XYZ anisotropy, spinons can move along one leg while creating virtual spinons on the other leg, implying that kinetic processes which are sensitive to the directionality of the spinon motion are present (see \ref{sec:semiclassical}).

One of the most interesting questions tackled in this study concerns the polarization of bound states. In the antiferromagnetic Ising spin chain there are two sequences of Zeeman ladders, composed of states with a total spin in the ordered direction of $S^\text{(total)}\!=\! 0$ and  $S^\text{(total)}\!=\!\pm 1$, contributing to the structure factors parallel and transverse to the ordered moment direction. 
In our model, due to general anisotropy, no projection of angular momentum is conserved, only the parity of the total $S^y$ spin. 
This results in distinct sequences of bound states in the longitudinal and transverse channels, as seen in the simulations, Fig.~\ref{fig:SSF_numerics}.
One of the most striking difference between the three spin directions is the gap to the lowest band, which is shifted to higher energies for $\mathcal{S}^{yy}$, Fig.~\ref{fig:SSF_numerics}(b).
This can be understood from the fact that the ground state of the Hamiltonian is N\'{e}el ordered in the $y$-direction. While the application of a $S^x$ or $S^z$ operator results in a spin flip and the overlap with the two spinon excitations, in the semiclassical limit the ground state is an eigenstate of the $S^y$ operators. Thus, $\mathcal{S}^{yy}$ connects to higher-order fluctuations present in the ground state resulting in a larger energy gap.
Interestingly, also the features stemming from frustration, the asymmetry with respect to $\mathbf{qb}=\pi$, and the overall intensity distribution differ between $\mathcal{S}^{xx}$, $\mathcal{S}^{yy}$, and $\mathcal{S}^{zz}$ [see Figs.~\ref{fig:SSF_numerics}(a)-(c)].

Previous INS experiments only hinted at the polarization dependence
of the bound states \cite{FacherisZheludev2023}. To fully resolve  this issue we performed a series of measurements on the THALES polarized-neutron 3-axis spectrometer at ILL \cite{datapolarized}. 
The same sample as used in Ref.~\cite{FacherisZheludev2023} was mounted with the crystallographic $\mathbf{b}$ and $\mathbf{c}$-axes in the scattering plane of the instrument. A polarized incident beam was produced by a Heussler monochromator. Combined with a Heussler analyzer, a flipping ratio of $R=17.6$ was achieved at a final neutron energy at 2.7~meV. This setup provided an energy resolution of $42~\mu$eV FWHM at the elastic line. The neutron beam was at all times polarized along the crystallographic $\mathbf{a}$ axis. All
data were collected at $T=50$~mK for a momentum transfer $\mb{q}=(0,0.5,3)$. 
In the non-spin-flip channel one measures excitations polarized along the $\mathbf{a}$ axis. Since the latter forms an angle of about 45$^\circ$ with the $x$ and $y$ axes \cite{PovarovZheludev2020, FacherisZheludev2024}, this measurement is sensitive to transverse excitations, namely to $\mathcal{S}^{xx}+\mathcal{S}^{zz}$ (see \ref{sec:INS_data}).
The spin-flip channel contains both this and the longitudinal contribution $\mathcal{S}^{yy}$. From the two measurements the longitudinal and transverse-polarized scattering are separated. The result is plotted in symbols in Figs.~\ref{fig:SSF_cuts}(c),(d). 

To within experimental resolution, the entirety of transverse-polarized scattering can be attributed to resolution-limited excitations.
Correspondingly, the data were analyzed by a model cross section composed of a series of sharp modes with parabolic dispersion relations along the $\mathbf{b}$ axis. The spin stiffness was fixed at the value previously determined using polarized neutrons \cite{FacherisZheludev2023}. 
The other parameters were the gap energies and intensities for each mode. This cross-section was numerically convoluted with the spectrometer resolution function computed in the Popovici approximation \cite{Popovici1975} using the ResLib library \cite{Zheludevreslib}. Results of this analysis are shown in solid lines in Figs.~\ref{fig:SSF_cuts}(c),(d).  
The positions of the first three peaks were fit, while those of the
others were fixed to values previously seen with unpolarized
spectroscopy \cite{FacherisZheludev2023}. The fitted excitation energies are plotted in Fig.~\ref{fig:SSF_cuts}. 
In the longitudinal channel there appears to be only a single resolution-limited peak [shaded area in Fig.~\ref{fig:SSF_cuts}(d), position also shown in Fig.~\ref{fig:SSF_cuts}(e)], followed by either a continuum of scattering or several poorly resolved modes.
In the numerical results we observe multiple peaks corresponding to the spinon bound states for both the longitudinal and transversal polarizations [Figs.~\ref{fig:SSF_cuts}(a),(b)].
The gap difference between the two polarizations and the energies of the bound states agree very well between the experimental and the numerical results of the triangular ladder [see Figs.~\ref{fig:SSF_cuts}(a),(b),(e)].

\section{Conclusions}

In conclusion, we investigated the confinement of spinon excitations in frustrated XYZ spin chains, by performing INS experiments on the compound Cs\textsubscript{2}CoBr\textsubscript{4} and numerically exact simulations of triangular spin-$1/2$ ladders.
From the excellent agreement of the experimental and theoretical results we identify coupled triangular ladders as the minimal model for describing the low-energy excitations of Cs\textsubscript{2}CoBr\textsubscript{4}. 
This allowed us to study the Zeeman ladder features appearing in the dynamical spin structure factor resulting from the spinon confinement in the presence of triangular couplings.
The hierarchical binding sequence of the spinon excitation is dominated by the one-dimensional physics of mean-field coupled XYZ chains, generalizing the previously observed instances in the regime of strong Ising anisotropy. 
However, the inclusion of the frustrated couplings of the triangular ladder is essential in recovering the dynamical response throughout the Brillouin zone. 
Furthermore, multiple Zeeman ladders are present for the different polarization channels with slight variations in the energy at $\mathbf{qb}=\pi$ and momentum dependence.

\section*{ACKNOWLEDGMENTS} 

This work was supported by the Swiss National Science Foundation under Division II grant 200020-219400 and partially supported by a MINT grant, SNSF Division 2.

\section*{DATA AVAILABILITY} 

The supporting data for the numerical simulations are openly available at Zenodo \cite{datazenodo}. The experimental data is available in Refs.~\cite{dataINS, datapolarized}.

\setcounter{section}{0}
\renewcommand{\thesection}{APPENDIX~\Alph{section}}
\renewcommand{\thesubsection}{\arabic{subsection}}
\renewcommand{\theequation}{A.\arabic{equation}}
\setcounter{equation}{0}

\section{\label{app_model} Connection to the spin-$3/2$ Hamiltonian}

In Ref.~\cite{FacherisZheludev2024} the experimentally measured excitations of the compound Cs\textsubscript{2}CoBr\textsubscript{4} were compared with results from spin wave theory of a spin-$3/2$ Hamiltonian which contained isotropic spin-spin couplings and large single ion anisotropies for the $x$ and $y$ directions.
In this work, we start from a similar spin-$3/2$ Hamiltonian, however we add the term breaking the symmetry between the $x$ and $y$ directions to the spin-spin couplings rather than the single ion anisotropies as in Ref.~\cite{FacherisZheludev2024}. The spin-$3/2$ Hamiltonian on a triangular ladder is given by
\begin{align}
\label{eq:Hamiltonian_32}
&\Tilde{H} =\\
&\Tilde{J}\sum_{j,m} \left[\Tilde{S}_{m,j}^x \Tilde{S}^x_{m,j+1}+ (1+\delta)\Tilde{S}_{m,j}^y \Tilde{S}^y_{m,j+1}+\Tilde{S}_{m,j}^z \Tilde{S}^z_{m,j+1}\right] \nonumber  \\
+&\Tilde{J}_1\sum_{j} \left[\Tilde{S}_{1,j}^x \Tilde{S}^x_{2,j}+ (1+\delta)\Tilde{S}_{1,j}^y \Tilde{S}^y_{2,j}+\Tilde{S}_{1,j}^z \Tilde{S}^z_{2,j}\right] \nonumber \\
+&\Tilde{J}_1\sum_{j} \left[\Tilde{S}_{1,j+1}^x \Tilde{S}^x_{2,j}+ (1+\delta)\Tilde{S}_{1,j+1}^y \Tilde{S}^y_{2,j}+\Tilde{S}_{1,j+1}^z \Tilde{S}^z_{2,j}\right] \nonumber  \\
-&D\sum_{j,m}\left[\left(\Tilde{S}_{m,j}^x\right)^2+ \left(\Tilde{S}_{m,j}^y\right)^2\right]. \nonumber 
\end{align}
As in this work we are interested in the low-energy excitation of the model, motivated by the large values of the single ion anisotropies $D$ in Cs\textsubscript{2}CoBr\textsubscript{4} \cite{FacherisZheludev2024} we derive a spin-$1/2$ Hamiltonian by restricting to the subspace spanned by the states $\ket{\Tilde{S^z}=\frac{1}{2}}$ and $\ket{\Tilde{S^z}=-\frac{1}{2}}$.
At the zeroth order in $1/D$ the projection to an effective spin $1/2$ model consists in replacing the spin $3/2$ operators $\Tilde{S}^\pm$ with the spin $1/2$ operators $2S^\pm$. This results in the following XYZ Hamiltonian 
\begin{align}
\label{eq:Hamiltonian_XYZ}
&H_{XYZ} = \\
&4\Tilde{J}\sum_{j,m} \big[ S_{m,j}^x S^x_{m,j+1}+(1+\delta) S_{m,j}^y S^y_{m,j+1}  \nonumber \\
&\qquad\qquad+0.25 S_{m,j}^z S^z_{m,j+1}\big] \nonumber \\
&+4\Tilde{J}_1\sum_{j} \big[S_{1,j}^x S^x_{2,j}+(1+\delta) S_{1,j}^y S^y_{2,j} +0.25 S_{1,j}^z S^z_{2,j} \nonumber  \\
&\qquad+S_{1,j+1}^x S^x_{2,j}+ (1+\delta)S_{1,j+1}^y S^y_{2,j} +0.25 S_{1,j+1}^z S^z_{2,j}\big]. \nonumber 
\end{align}
Which is the same Hamiltonian as introduced in the main text with $J\equiv 4\Tilde{J}$, $J_1\equiv 4\Tilde{J_1}$ and $\Delta_z=0.25$.
We note that in contrast to the Hamiltonian discussed in Ref.~\cite{FacherisZheludev2024}, the one we consider here in Eq.~(\ref{eq:Hamiltonian_32}) leads to a spin-$1/2$ Hamiltonian for which the symmetry between the $x$ and $y$ directions is broken already at the zeroth order in $1/D$, as seen in Eq.~(\ref{eq:Hamiltonian_XYZ}).

\setcounter{equation}{0}
\renewcommand{\theequation}{B.\arabic{equation}}
\setcounter{figure}{0}
\renewcommand{\thefigure}{B\arabic{figure}}

\section{\label{app_methods} Theoretical methods}

\subsection{Numerical Matrix Product States methods \label{app_mps}}

The numerical results for the dynamical correlations presented in this work are obtained using a time-dependent matrix product state (tMPS) method.
Our implementation is based on the Trotter-Suzuki decomposition of the time evolution operator \cite{WhiteFeiguin2004, DaleyVidal2004, Schollwoeck2011} and makes use of the ITensor Library \cite{FishmanStoudenmire2020}. 
The ground state of the model has been determined using the density matrix renormalization group (DMRG) algorithm in the matrix product state (MPS) representation \cite{White1992, Schollwoeck2011}. 
For the dynamical spin structure factor presented we consider systems of $L=128$ sites and take a maximal bond dimension up to 350 states, which ensures that the truncation error is at most $5\times 10^{-8}$ at the final time of the evolution, $tJ/\hbar=62.5$. 
For the parameters employed in the comparison with the experimental results we checked that a maximal bond dimension up to 450 states, which ensures that the truncation error is at most $5\times 10^{-9}$, does not produce any visible effect of the computed dynamical correlations. 
We employ at time step of $dtJ/\hbar=0.02$ and measure the observables every fifth time step.

\subsection{Analytical Bosonization approach}

We can describe the low-energy physics of one-dimensional interacting quantum systems corresponding to the Tomonaga-Luttinger liquid universality class in terms of two bosonic fields $\phi$ and $\theta$ \cite{Giamarchibook}.
The bosonic fields fulfill the canonical commutation relation, $\left[ \phi(x),\nabla\theta(x') \right]=i\pi\delta(x-x')$.
In the bosonized representation, in the absence of magnetization, the spin operators for a single chain can be written as \cite{Giamarchibook}
\begin{align} 
\label{eq:bosonization}
S^+_j&=\frac{1}{\sqrt{2\pi}}e^{-i\theta(x)}\left[(-1)^x+\cos(2\phi(x))\right]\\
S^z_j&=-\frac{1}{\pi}\partial_x \phi(x)+\frac{(-1)^x}{\pi}\cos(2\phi(x)).\nonumber
\end{align}
where the coordinate is $x=ja$, with $a$ the lattice spacing which we fix to $a=1$ in the following.

We apply the bosonization procedure for the Hamiltonian $H=H_{XYZ}+H_{MF}$ (as defined in the main text) in the limit of vanishing $\Delta_z$.
We first perform the following rotation of the spin operators basis
\begin{align}
\label{eq:rotation}
S^x_{m,j} &\to -S^y_{m,j}, \\ 
S^y_{m,j} &\to S^x_{m,j}, \nonumber \\
S^z_{m,j} &\to S^z_{m,j}, \nonumber
\end{align}
followed by the rewriting of the spin operators in terms of $S^+_{m,j}$ and $S^-_{m,j}$ operators to obtain
\begin{align}
\label{eq:Hamiltonian_XY}
H(\Delta_z=&0) =\\
J\sum_{j,m} &\Bigg[\frac{2+\delta}{4} \left(S_{m,j}^+ S^-_{m,j+1}+ S_{m,j}^- S^+_{m,j+1}\right)  \nonumber \\
&+\frac{\delta}{4} \left(S_{m,j}^+ S^+_{m,j+1}+ S_{m,j}^- S^-_{m,j+1}\right)\Bigg] \nonumber \\
+J_1\sum_{j}&\Bigg[\frac{2+\delta}{4}  \left(S_{1,j}^+ S^-_{2,j}+ S_{1,j}^- S^+_{2,j}\right) \nonumber  \\
&+\frac{\delta}{4} \left(S_{1,j}^+ S^+_{2,j}+ S_{1,j}^- S^-_{2,j}\right) \nonumber  \\
&+\frac{2+\delta}{4}  \left(S_{1,j+1}^+ S^-_{2,j}+ S_{1,j+1}^- S^+_{2,j}\right) \nonumber  \\
&+\frac{\delta}{4} \left(S_{1,j+1}^+ S^+_{2,j}+ S_{1,j+1}^- S^-_{2,j}\right)\Bigg] \nonumber  \\
-\frac{J_3}{2} \sum_{j,m} & (-1)^{j+m}\left(S^+_{m,j}+S^-_{m,j}\right).\nonumber
\end{align}
We bosonize the Hamiltonian of the two coupled chains, Eq.~(\ref{eq:Hamiltonian_XY}), in the limit where coupling $J$ along the two chains dominates. In this regime we have a pair of bosonic fields for each leg of the ladder. By only considering that $S^+_{m,j}\propto e^{i\theta_m(x)} (-1)^j$ in the bosonization formula and that $x=aj$ for $m=1$ and $x=aj+1/2$ for $m=2$, we obtain
\begin{align} 
\label{eq:Hamiltonian_2chains_boson}
 H_\text{2~chains}= & \int\frac{dx}{2\pi}\left[uK\partial_x\theta_1^2+\frac{u}{K}\partial_x\phi_1^2\right] \\
 +& \int\frac{dx}{2\pi}\left[uK\partial_x\theta_2^2+\frac{u}{K}\partial_x\phi_2^2\right] \nonumber \\
 \label{eq:Hamiltonian_2chains_boson_2}
 -&g\delta\int dx\left[\cos(2\theta_1)+\cos(2\theta_2)\right]  \\
  \label{eq:Hamiltonian_2chains_boson_3}
 +&g_1(2+\delta)\int dx\sin\left[\theta_1-\theta_2+\frac{a}{4}\partial_x(\theta_1-\theta_2)\right] \nonumber \\
 &\qquad\qquad\quad\times\sin\left[\frac{a}{4}\partial_x(\theta_1+\theta_2)\right] \\
  \label{eq:Hamiltonian_2chains_boson_4}
 +&g_1\delta\int dx\sin\left[\theta_1+\theta_2+\frac{a}{4}\partial_x(\theta_1+\theta_2)\right] \nonumber \\
 &\qquad\quad\times\sin\left[\frac{a}{4}\partial_x(\theta_1-\theta_2)\right]  \\
\label{eq:Hamiltonian_2chains_boson_5}
 +&g_3\int dx\left[\cos(\theta_1)-\cos(\theta_2)\right],
\end{align}
with $K$ the Luttinger parameter and $u$ the velocity, and the coupling constants $g\propto J$, $g_1\propto J_1$, $g_3\propto J_3$.
The first two lines correspond to the quadratic part of the Hamiltonian for each leg, the term proportional to $g\delta$ stems from the $\left(S_{m,j}^+ S^+_{m,j+1}+\text{H.c.}\right)$ couplings along the legs [Eq.~(\ref{eq:Hamiltonian_2chains_boson_2})], the rung coupling gives the terms proportional to $g_1(2+\delta)$ [Eq.~(\ref{eq:Hamiltonian_2chains_boson_3})] for the $\left(S_{1,j}^+ S^-_{2,j}+S_{1,j+1}^+ S^-_{2,j}+\text{H.c.}\right)$ processes and $g_1\delta$ [Eq.~(\ref{eq:Hamiltonian_2chains_boson_4})] for the $\left(S_{1,j}^+ S^+_{2,j}+S_{1,j+1}^+ S^+_{2,j}+\text{H.c.}\right)$ processes, the $g_3$ stems from the effective staggered magnetic field.
We note that the signs of the terms in Eq.~(\ref{eq:Hamiltonian_2chains_boson_5}) fix the obtained spin configuration.
We further discuss the physics of the bosonized Hamiltonian, Eq.~(\ref{eq:Hamiltonian_2chains_boson}), together with the numerical results for the ground state in \ref{sec:app_gs}.

\setcounter{equation}{0}
\renewcommand{\theequation}{C.\arabic{equation}}
\setcounter{figure}{0}
\renewcommand{\thefigure}{C\arabic{figure}}
\section{Ground state calculations\label{sec:app_gs}}

In this appendix, we present numerical results for the ground state of the model, which we discuss together with insight obtain from the bosonization approach.
We note that in this section we employ the spin basis in which the following rotation has been performed compared to the results shown in the main text of the paper,
\begin{align}
\label{eq:rotation2}
S^x_{m,j} &\to -S^y_{m,j}, \\ 
S^y_{m,j} &\to S^x_{m,j}, \nonumber \\
S^z_{m,j} &\to S^z_{m,j}, \nonumber
\end{align}

\subsection{Mean-field treatment of the interladder coupling}

\begin{figure}[!hbtp]
\centering
\includegraphics[width=0.38\textwidth]{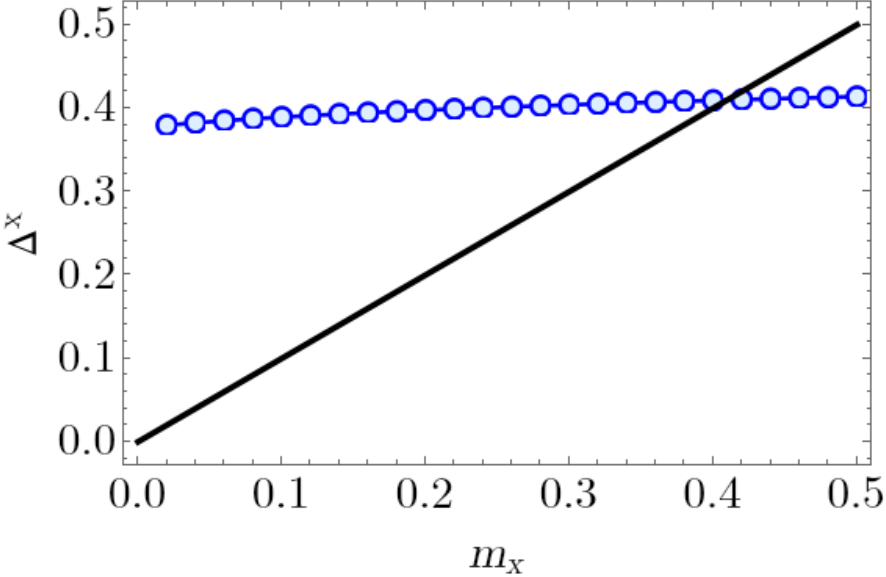}
\caption{\label{fig:mf_inter}
The staggered magnetization $\Delta^x$ as a function of the mean-field $m_x$ in the ground state of the model, used in the self-consistent determination of the strength of the interladder coupling. The solution is given by the intersection of the numerical curve with the black straight line at $m_x\approx 0.4$, determining a value of $J_3/J=0.05$.
Note that the spin basis differs by the rotation given in Eq.~(\ref{eq:rotation2}) compared to the main text.
The numerical DMRG ground state results are obtained for the parameters $L=160$ sites, $J_1/J=0.45$, $\delta=0.45$, $J_\text{inter}/J=0.125$, $\Delta_z=0.25$.}
\end{figure}

In this appendix, we briefly discuss the self-consistent mean-field approach which describes the interladder coupling, $H_{MF}=-J_\text{inter}\sum_{m,j} \langle S_{m,j}^x \rangle S^x_{m,j}$.
In the presence of a finite $xy$ anisotropy $\delta$, the ground state of the model [Eq.~(1) and Eq.~(2) in the main text] exhibits a N\'eel ordering along the chains of the ladder. 
Thus, we can take the expectation value of the magnetization in the $x$-direction to be described by the mean-field $\langle S_{m,j}^x \rangle\approx (-1)^{j+m} m_x$.
To determine the value of $m_x$ we compute the value of the staggered magnetization $\Delta^x=\frac{1}{L}\sum_{j,m} (-1)^{j+m} \langle  S_{m,j}^x \rangle$ for different $m_x$, as depicted in Fig.~\ref{fig:mf_inter}. The self-consistently determined value of $m_x$ is fixed by the intersection of the two curves.
In the following, we define $J_3\equiv J_\text{inter} m_x$.

\subsection{Nature of the ground state}

\begin{figure}[!hbtp]
\centering
\includegraphics[width=0.38\textwidth]{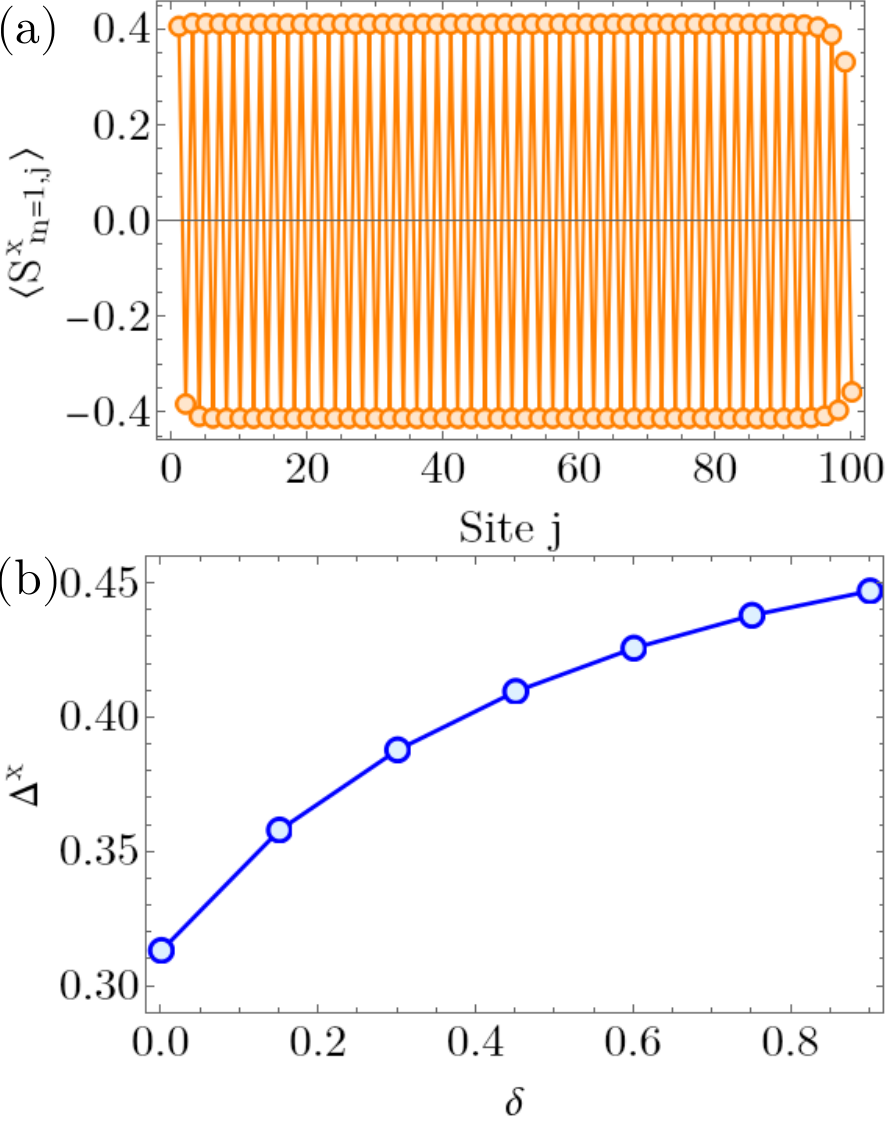}
\caption{\label{fig:stag_dz}
(a) The spatial dependence of the local magnetization $\langle S^x_{m=1,j}\rangle$ along a leg, for $\delta=0.45$. 
(b) The staggered magnetization $\Delta^x$ as a function of $\delta$.
Note that the spin basis differs by the rotation given in Eq.~(\ref{eq:rotation2}) compared to the main text.
The numerical DMRG ground state results are obtained for the parameters $L=200$ sites, $J_1/J=0.45$, $J_\text{inter}/J=0.125$, $\Delta_z=0.25$.
In panel (a) the self-consistent value of the interladder coupling is $J_3/J=0.05$.}
\end{figure}

\begin{figure}[!hbtp]
\centering
\includegraphics[width=0.38\textwidth]{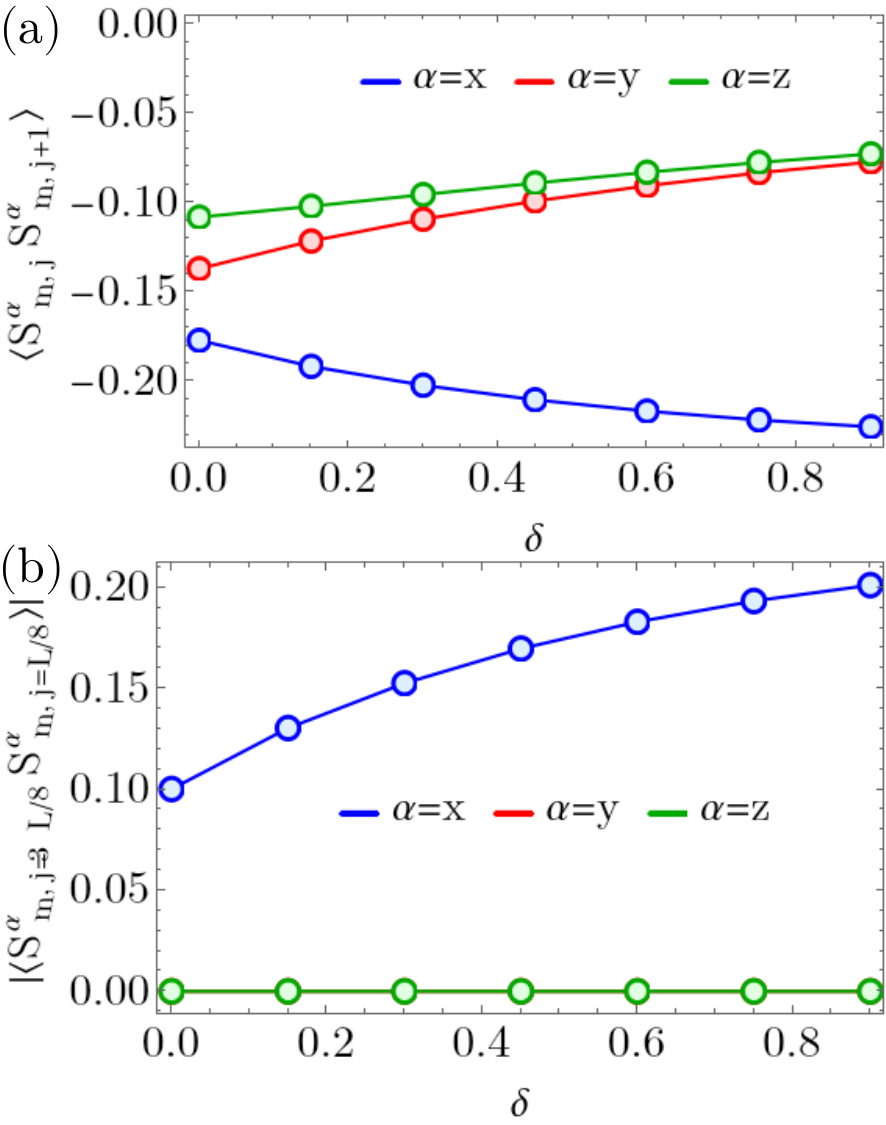}
\caption{\label{fig:corr_leg}
(a) The spin-spin correlation function of neighboring spins $\langle S^\alpha_{m=1,j}S^\alpha_{m=1,j+1}\rangle$, averaged along a leg, as a function of $\delta$, for different spin directions $\alpha\in\{x,y,z\}$. 
(b) The long-distance spin-spin correlation function $\langle S^\alpha_{m=1,j=3L/8}S^\alpha_{m=1,j=L/8}\rangle$ as a function of $\delta$, for different spin directions $\alpha\in\{x,y,z\}$. 
Note that the spin basis differs by the rotation given in Eq.~(\ref{eq:rotation2}) compared to the main text.
The numerical DMRG ground state results are obtained for the parameters $L=200$ sites, $J_1/J=0.45$, $J_\text{inter}/J=0.125$, $\Delta_z=0.25$.}
\end{figure}

\begin{figure}[!hbtp]
\centering
\includegraphics[width=0.49\textwidth]{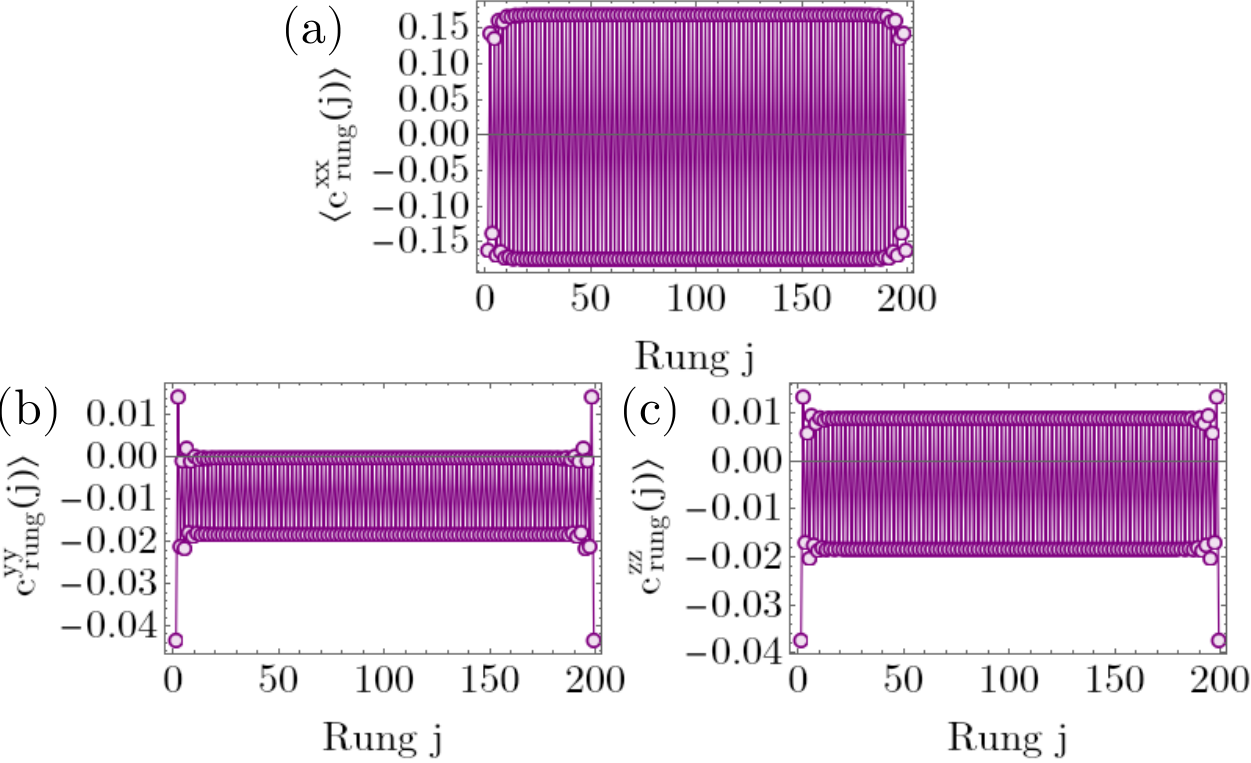}
\caption{\label{fig:corr_rung}
The spin-spin rung correlation function $c^{\alpha\alpha}_\text{rung}$, defined in Eq.~(\ref{eq:rung_corr}), as a function of position, for different spin directions (a) $\alpha=x$, (b) $\alpha=y$, (c) $\alpha=z$. 
Note that the spin basis differs by the rotation given in Eq.~(\ref{eq:rotation2}) compared to the main text.
The numerical DMRG ground state results are obtained for the parameters $L=200$ sites, $\delta=0.45$, $J_1/J=0.45$, $J_\text{inter}/J=0.125$, $J_3/J=0.05$, $\Delta_z=0.25$.}
\end{figure}

In this appendix, we analyze the numerical results for the ground state of the Hamiltonian $H=H_{XYZ}+H_{MF}$.
Both the anisotropy $\delta$ and the effective magnetic field stemming from the interladder coupling break the symmetry between the $x$ and $y$ directions and induce a finite gap. 
We note that in the absence of $\delta$ and $J_\text{inter}$ for the considered parameters the underlying XXZ model on the triangular ladder would result in a gapless state.
We observe in Fig.~\ref{fig:stag_dz}(a) that for finite values of $\delta$ and $J_\text{inter}$ the magnetization in the $x$ direction along the leg of the ladder exhibits an N\'eel pattern with a finite value of the staggered magnetization $\Delta^x=\frac{1}{L}\sum_{j,m} (-1)^{j+m} \langle  S_{m,j}^x \rangle$ [as depicted in Fig.~\ref{fig:stag_dz}(b)].
We also checked the gapped nature of the state by computing the scaling of the von Neumman entanglement entropy in the system (not shown) and obtained a central charge consistent with zero.
Such a phase has been obtained also for $J_1$-$J_2$ XXZ spin chains and called the UUDD phase, for example see \cite{Igarashi1989, FurukawaFurusaki2012}. 

Furthermore, the correlations between neighboring spins along the legs of the ladder have an antiferromagnetic character, as seen in Fig.~\ref{fig:corr_leg}(a), where $\langle S^\alpha_{m=1,j}S^\alpha_{m=1,j+1}\rangle$ are negative for all spin directions.
While in the $y$ and $z$ directions we do not obtain any local magnetizations, we still have finite antiferromagnetic correlations between neighboring spins. However, in Fig.~\ref{fig:corr_leg}(b), where we show the correlations between spins at a distance of $L/4$ we can see that the long-range order is only present in the $x$ direction.

In order to investigate the coupling of the spins along the rungs of the ladder we define the following correlation function
\begin{align}
\label{eq:rung_corr}
c^{\alpha\alpha}_\text{rung}(2j-1)&=S^\alpha_{j,1} S^\alpha_{j,2}, \\
c^{\alpha\alpha}_\text{rung}(2j)&=S^\alpha_{j+1,1} S^\alpha_{j,2}. \nonumber
\end{align}
We show the ground state results in Fig.~\ref{fig:corr_rung}. From the behavior of $c^{xx}_\text{rung}$ [see Fig.~\ref{fig:corr_rung}(a)] we can infer that on a triangular plaquette one of the rungs the correlations are antiferromagnetic, while on the other rung they have a ferromagnetic character.
This supports the conclusion that one each leg of the ladder the spins are ordered in a N\'eel pattern.
The correlations in the other directions,  Fig.~\ref{fig:corr_rung}(b)-(c), are weaker, and while they exhibit an alternate pattern their average value is non-zero. This stems from the frustration effects appearing beyond the semiclassical order, as we discuss in the following appendix.

\begin{figure}[!hbtp]
\centering
\includegraphics[width=0.38\textwidth]{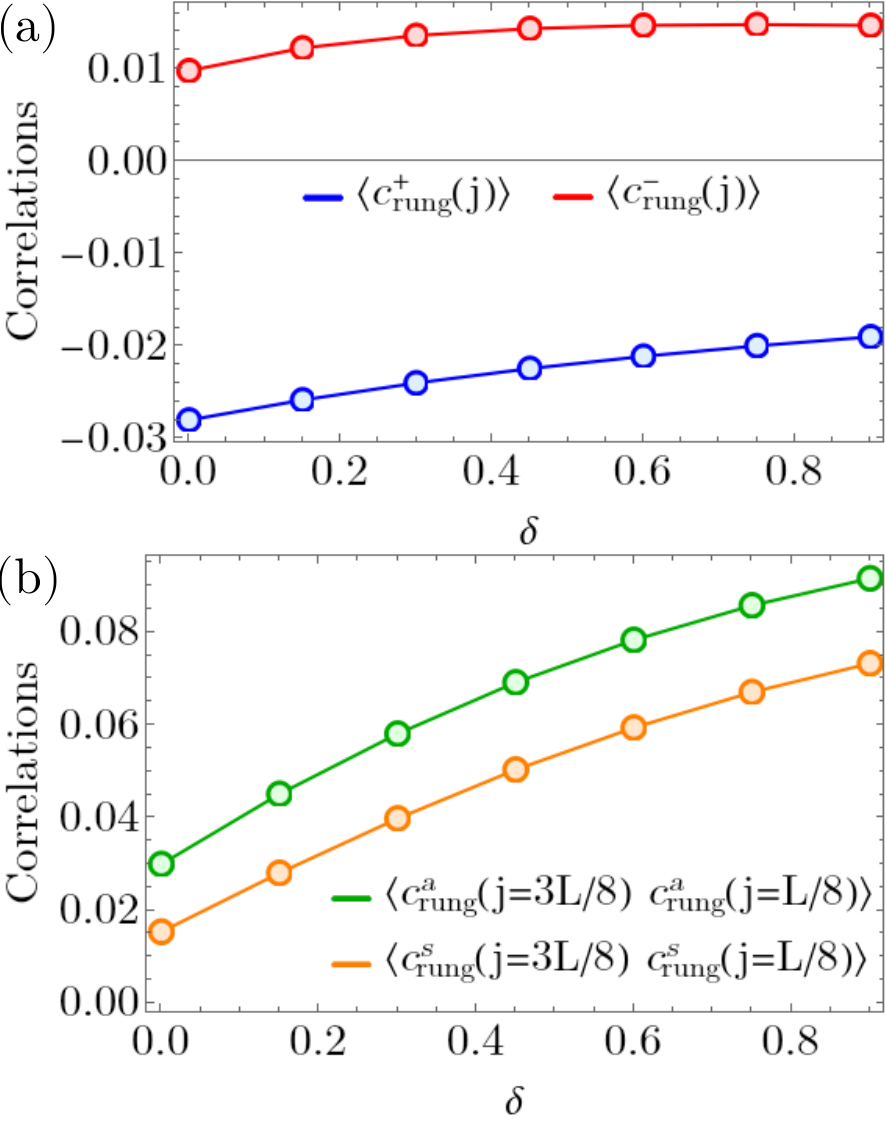}
\caption{\label{fig:corr_rung_as}
The spin-spin rung correlation function, defined in (a) Eq.~(\ref{eq:rung_corr_2}) and (b) Eq.~(\ref{eq:rung_corr_3}). 
Note that the spin basis differs by the rotation given in Eq.~(\ref{eq:rotation2}) compared to the main text.
The numerical DMRG ground state results are obtained for the parameters $L=200$ sites, $J_1/J=0.45$, $J_\text{inter}/J=0.125$, $J_3/J=0.05$, $\Delta_z=0.25$.}
\end{figure}

\subsection{Role of the frustrated geometry \label{app_frustration}}

In the previous appendix we have seen that the numerical results for the ground state of the model for the considered parameters exhibits  antiferromagnetic order along each leg of the ladder.
At a semiclassical level such order would determine that the two legs are decoupled due to the triangular geometry, as each spin would be coupled to two spin from the other leg, each pointing in opposite directions, resulting in average to a vanishing coupling.
We can understand this also from the bosonized Hamiltonian, Eqs.~(\ref{eq:Hamiltonian_2chains_boson})-(\ref{eq:Hamiltonian_2chains_boson_5}).
The antiferromagnetic order stems from the terms breaking the $xy$ symmetry, meaning that the terms from Eq.~(\ref{eq:Hamiltonian_2chains_boson_2}) and Eq.~(\ref{eq:Hamiltonian_2chains_boson_5}) are the relevant ones in the bosonized Hamiltonian.
In a semiclassical analysis it results the $\cos$ potentials will fix the bosonic fields to the values $\theta_1=\pi$ and $\theta_2=0$, minimizing the value of the energy stemming from the terms Eq.~(\ref{eq:Hamiltonian_2chains_boson_2}) and Eq.~(\ref{eq:Hamiltonian_2chains_boson_5}).
For these values of the fields the terms corresponding to the triangular coupling between the legs in Eq.~(\ref{eq:Hamiltonian_2chains_boson_3}) and Eq.~(\ref{eq:Hamiltonian_2chains_boson_4}) vanish.
This implies that the dominating physics is that of decoupled single chains. However, as we will see in the following we can identify certain correlations that would vanish in the semiclassical analysis, i.e.~for $\theta_1=\pi$ and $\theta_2=0$, but have finite values in the numerically computed ground state of the model.

We first define the following correlations, which correspond to the local expectation values of the terms in Eq.~(\ref{eq:Hamiltonian_2chains_boson_3}) and Eq.~(\ref{eq:Hamiltonian_2chains_boson_4}),
\begin{align}
\label{eq:rung_corr_2}
c^+_\text{rung}&(j)= \\
=&c^{xx}_\text{rung}(2j-1)+c^{yy}_\text{rung}(2j-1)+c^{xx}_\text{rung}(2j)+c^{yy}_\text{rung}(2j)\nonumber \\ 
=&S_{m,j}^+ S^-_{m,j+1}+ S_{m,j}^- S^+_{m,j+1} +S_{1,j+1}^+ S^-_{2,j}+ S_{1,j+1}^- S^+_{2,j} \nonumber\\ 
\propto&\sin\left[\theta_1-\theta_2+\frac{a}{4}\partial_x(\theta_1-\theta_2)\right]\sin\left[\frac{a}{4}\partial_x(\theta_1+\theta_2)\right], \nonumber \\ 
c^-_\text{rung}&(j)=  \nonumber\\
=&c^{xx}_\text{rung}(2j-1)-c^{yy}_\text{rung}(2j-1)+c^{xx}_\text{rung}(2j)-c^{yy}_\text{rung}(2j)\nonumber \\ 
=&S_{m,j}^+ S^+_{m,j+1}+ S_{m,j}^- S^-_{m,j+1} +S_{1,j+1}^+ S^+_{2,j}+ S_{1,j+1}^- S^-_{2,j} \nonumber\\ 
\propto&\sin\left[\theta_1+\theta_2+\frac{a}{4}\partial_x(\theta_1+\theta_2)\right]\sin\left[\frac{a}{4}\partial_x(\theta_1-\theta_2)\right]. \nonumber 
\end{align}
In Fig.~\ref{fig:corr_rung_as}(a) we observe that the expectation values of the correlations $c^+_\text{rung}$ and $c^-_\text{rung}$, averaged over the entire system, have finite values.
The finite values imply that we need to have deviations from the $\theta_1=\pi$ and $\theta_2=0$ values of the fields.
In order to further probe this we investigate the long-range behavior of the correlations given by the antisymmetric and symmetric combinations of the two bosonic fields
\begin{align}
\label{eq:rung_corr_3}
c^a_\text{rung}&(j)=S_{m,j}^+ S^-_{m,j+1} - S_{m,j}^- S^+_{m,j+1} \\ 
&\propto\sin\left[\theta_1-\theta_2-\frac{a}{4}\partial_x\theta_2\right], \nonumber \\ 
c^s_\text{rung}&(j)=S_{m,j}^+ S^+_{m,j+1} - S_{m,j}^- S^-_{m,j+1} \nonumber \\ 
&\propto\sin\left[\theta_1+\theta_2-\frac{a}{4}\partial_x\theta_2\right]. \nonumber 
\end{align}
The expectation values $\langle c^{a/s}_\text{rung}(L/8+d)c^{a/s}_\text{rung}(L/8) \rangle$ as a function of the distance $d$ show a rapid exponential decay to a finite value, behavior characteristic to the gapped phase. In Fig.~\ref{fig:corr_rung_as}(b) we show the dependence of the value at which the long-range correlations saturate as a function of $\delta$.
From this we can see that the deviations from the $\theta_1=\pi$ and $\theta_2=0$ values of the fields do not characterize only the local behavior of correlations [as seen in Fig.~\ref{fig:corr_rung_as}(a)], but also the long-range order which stabilizes in the system [see Fig.~\ref{fig:corr_rung_as}(b)].
Thus, we can conclude that in the numerical exact results for the ground state of the model we have identified correlations describing the frustrated coupling between the legs of the ladder, which go beyond the semiclassical description of the N\'eel order.
Furthermore, we note that the terms describing the triangular coupling, Eq.~(\ref{eq:Hamiltonian_2chains_boson_3}) and Eq.~(\ref{eq:Hamiltonian_2chains_boson_4}), are very different from the terms that would be obtained by coupling the two legs in a square geometry.

\setcounter{equation}{0}
\renewcommand{\theequation}{D.\arabic{equation}}
\setcounter{figure}{0}
\renewcommand{\thefigure}{D\arabic{figure}}

\section{Semiclassical limit and quantum fluctuations\label{sec:semiclassical}}

In this appendix, we discuss the semiclassical limit of the model and the quantum processes involving the spinon bound states which are sensitive to the triangular geometry.
In order to facilitate the discussion we perform the following spin rotations, mapping the N\'eel order in the $y$-direction, discussed in the main text, to a fully polarized state in the $z$-direction
\begin{align}
\label{eq:rotation3}
S^x_{m,j} &\to -S^x_{m,j} \to -S^x_{m,j}, \\ 
S^y_{m,j} &\to S^z_{m,j} \to (-1)^{j+m} S^z_{m,j}, \nonumber \\
S^z_{m,j} &\to S^y_{m,j} \to (-1)^{j+m} S^y_{m,j}. \nonumber
\end{align}
This transformation leads to the following Hamiltonian in terms of $S^+_{m,j}$ and $S^-_{m,j}$  operators 
\begin{align}
\label{eq:Hamiltonian_rotated1}
H=-J\sum_{j,m} &\Bigg[(1+\delta)S_{m,j}^z S^z_{m,j+1}  \\
-&\frac{1+\Delta_z}{4} \left(S_{m,j}^+ S^+_{m,j+1}+S_{m,j}^- S^-_{m,j+1} \right) \nonumber\\
-&\frac{1-\Delta_z}{4}  \left(S_{m,j}^+ S^-_{m,j+1}+ S_{m,j}^- S^+_{m,j+1}\right)\Bigg] \nonumber \\
-J_1\sum_{j}&\Bigg[(1+\delta)S_{1,j}^z S^z_{2,j}\nonumber\\
-&\frac{1+\Delta_z}{4} \left(S_{1,j}^+ S^+_{2,j}+S_{1,j}^- S^-_{2,j} \right)   \nonumber\\
-&\frac{1-\Delta_z}{4}  \left(S_{1,j}^+ S^-_{2,j}+ S_{1,j}^- S^+_{2,j}\right)\Bigg] \nonumber \\
+J_1\sum_{j}&\Bigg[(1+\delta)S_{1,j+1}^z S^z_{2,j}   \nonumber\\
-&\frac{1-\Delta_z}{4} \left(S_{1,j+1}^+ S^+_{2,j}+S_{1,j+1}^- S^-_{2,j} \right) \nonumber\\
-&\frac{1+\Delta_z}{4}  \left(S_{1,j+1}^+ S^-_{2,j}+ S_{1,j+1}^- S^+_{2,j}\right)\Bigg] \nonumber \\
-h \sum_{j,m} & S^z_{m,j}.\nonumber
\end{align}

 \begin{figure}[!hbtp]
 \centering
 \includegraphics[width=0.4\textwidth]{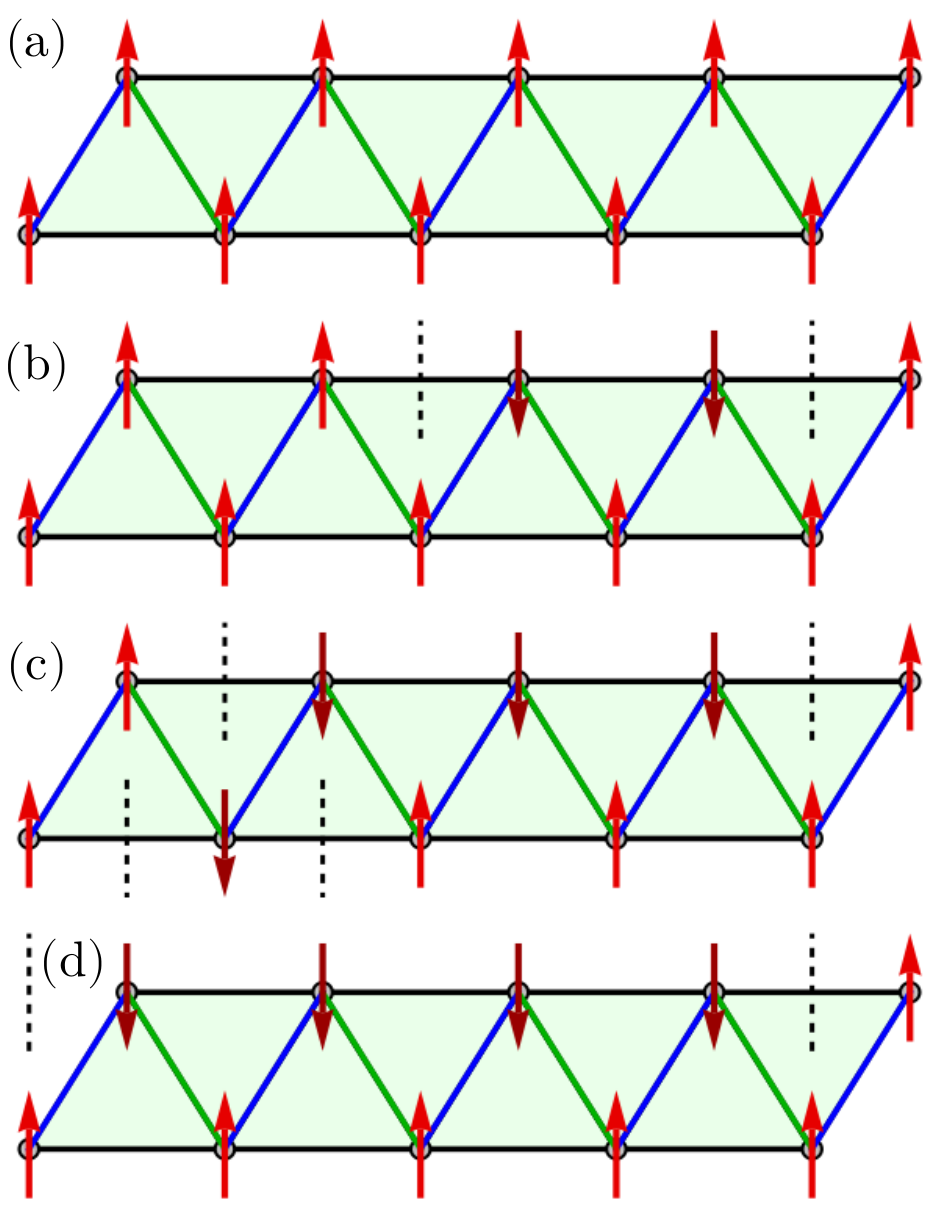}
 \caption{\label{fig:semiclassical_pattern}
Semiclassical spin patterns corresponding approximating the (a) ground state of the Hamiltonian in Eq.~(\ref{eq:Hamiltonian_rotated1}) with a fully polarized state in the $z$-direction and (b) a two spinon bound state located on a leg of the ladder, between the black dashed lines.
(c)-(d) Examples of spin-flip processes leading to the dynamics of the spinon bound state, see text after Eq.~(\ref{eq:Hamiltonian_rotated1}) for the corresponding discussion. 
The alternation of blue and green rungs corresponds to the different signs of the couplings on the respective rungs as given in Eq.~(\ref{eq:Hamiltonian_rotated1}).}
 \end{figure}

In the rotated spin basis of the Hamiltonian in Eq.~(\ref{eq:Hamiltonian_rotated1}) the ground state N\'eel order discussed in previous section corresponds to a fully polarized state in the $z$-direction. The semiclassical spin pattern of the ground state, in the absence of quantum correlations, is depicted in Fig.~\ref{fig:semiclassical_pattern}(a).
We can also observe in Eq.~(\ref{eq:Hamiltonian_rotated1}) that the couplings in the $z$-direction of the triangular rungs have alternating signs (shown with blue and green color in Fig.~\ref{fig:semiclassical_pattern}). This results in the fact that at the semiclassical level a spin is coupled with two ordered spins from the other leg with equal and opposite sign couplings, effectively resulting in a decoupling of the two legs of the triangular ladder.

Within the semiclassical approximation a spinon bound states corresponds to a string of flipped spins on one leg, as shown in Fig.~\ref{fig:semiclassical_pattern}(b) between the vertical dashed lines. 
The energy of this spin pattern compared to the ground state is $J(1+\delta)$, plus the contribution from the magnetic field related to the length of the string. 
If we look at the local potential energies on the second leg, we can see that the spins corresponding to the beginning and the end of the string of flipped spins [the spins at the position of the domain walls shown with the vertical dashed lines in Fig.~\ref{fig:semiclassical_pattern}(b)], experience a local potential of $\pm J_1(1+\delta)$.
As in our case $J_1<J$ this change in potential energy is not enough to flip the spins on the second leg, but it provides a first evidence that for the states relevant for our discussion the spins on one leg can be sensitive to the spin pattern on the other leg.

The potential energy induced by the spinon bound state on the other leg is attractive at one end of the string of flipped spins and repulsive at the other, thus, it can lead to the momentum asymmetry observed in the dynamical response discussed in the main text.
However, in order for this effect to be relevant we need to go beyond the semiclassical limit and include quantum fluctuations that can create virtual spinons on the second leg on the sites with a modified energy.
In particular, one possible process leading to such quantum fluctuations would be that starting from the state from Fig.~\ref{fig:semiclassical_pattern}(b) we act with an $S^-S^-$ term on the blue rung to increase the length of the spinon bound state, while creating an additional spin flip on the other leg on a site with a modified energy [see Fig.~\ref{fig:semiclassical_pattern}(c)].
This is followed by the action of an $S^-S^+$ term on the green rung, leading to a longer spinon bound state on the first leg and back to a fully polarized spin pattern on the second leg, as shown in Fig.~\ref{fig:semiclassical_pattern}(d).
This second order process would lead to a dynamics of the spinon bound states which is sensitive to the direction in which the domain walls propagate, and can determine the momentum asymmetry discussed in the main text.
We emphasize that the anisotropy structure of our model is essential for the presence of the quantum fluctuations that ensure the effective coupling between the two legs of the triangular ladder and lead to the dynamical response observed in the numerical and experimental results.

\setcounter{equation}{0}
\renewcommand{\theequation}{E.\arabic{equation}}
\setcounter{figure}{0}
\renewcommand{\thefigure}{E\arabic{figure}}
\section{Dynamical spin structure factor\label{sec:app_SSF}}

The experimental neutron scattering signal is given by  \cite{Lovesey1984, Squires2012}
\begin{align}
\label{eq:intensity}
I (\mathbf{q},\omega)= F(q)^2 \sum_\alpha
\left(1-\frac{q_\alpha^2}{q^2}\right)\mathcal{S}^{\alpha\alpha}(\mathbf{q},\omega),
\end{align}
where $\mathcal{S}^{\alpha\alpha}$ is the dynamical spin structure factor, $F(q)$ is the magnetic form factor, which we will discuss at the end of this section, and $q=\sqrt{\mathbf{q}^2}$.

The dynamical spin structure factor is defined as
\begin{align}
\label{eq:structure_factor}
\mathcal{S}^{\alpha\beta}(\mathbf{q}, \omega)=\frac{1}{L}\sum_{j,j'}
\int_{-\infty}^{\infty} dt\, e^{i \omega t} \langle S_{j'}^\alpha(0)
S_{j}^\beta(t) \rangle  e^{i \mathbf{q}\cdot
(\mathbf{r}_j-\mathbf{r}_{j'})}.
\end{align}
We compute the spin structure factor from tMPS simulations (see \ref{app_mps}). In the following we present the steps we took to perform the numerical calculations.

In the numerical implementation, we map the ladder Hamiltonian to the chain representation given by the Hamiltonian
\begin{align}
\label{eq:Hamiltonian_chain}
H_\text{chain}& =-J_3\sum_{j} h_j S_{j}^y \\
&+J\sum_{j} \left[S_{j}^x S^x_{j+2}+ (1+\delta)S_{j}^y S^y_{j+2} +\Delta_z S_{j}^z S^z_{j+2}\right] \nonumber \\
&+J_1\sum_{j} \left[S_{j}^x S^x_{j+1}+ (1+\delta)S_{j}^y S^y_{j+1} +\Delta_z S_{j}^z S^z_{j+1}\right] \nonumber, 
\end{align}
with $h_{4j+1}=1,~h_{4j+2}=-1,~h_{4j+3}=-1,~\text{and}~h_{4j+4}=1$.

We make use of translational invariance to restrict the sum over $j'$ to the distinct sites of the unit cell of the ladder only, which up to finite size effects recovers the structure factor from Eq.~(\ref{eq:structure_factor}). Thus, we take $j'\in\{L/2,L/2+1\}$ and obtain
\begin{align}
\mathcal{S}^{\alpha\alpha}(\mathbf{q}, \omega)=&\frac{1}{\sqrt{2L}}\sum_{j}\int_{-\infty}^{\infty} dt\, e^{i \omega t} \\
\times&\Big[\langle S_{L/2}^\alpha(0) S_{j}^\alpha(t) \rangle  e^{-i \mathbf{q}\cdot(\mathbf{r}_{L/2}-\mathbf{r}_{j})}\nonumber\\
&+\langle S_{L/2+1}^\alpha(0)S_{j}^\alpha(t) \rangle  e^{-i \mathbf{q}\cdot(\mathbf{r}_{L/2+1}-\mathbf{r}_{j})}\Big].\nonumber
\end{align}
For the finite size system considered in the simulations we can map one chain of the ladder into the other and arrive at the following relation between the spin-spin correlations
\begin{align}
\langle S_{L/2+1}^\alpha(0)S_{j}^\alpha(t) \rangle =\langle S_{L/2}^\alpha(0)S_{L+1-j}^\alpha(t) \rangle.
\end{align}
This leads to the following form of the structure factor
\begin{align}
\label{eq:SF2}
\mathcal{S}^{\alpha\alpha}(\mathbf{q}, \omega)=&\frac{1}{\sqrt{2L}}\sum_{j}\int_{-\infty}^{\infty} dt\, e^{i \omega t} \\
\times&\Big[\langle S_{L/2+1}^\alpha(0) S_{L+1-j}^\alpha(t) \rangle  e^{-i \mathbf{q}\cdot(\mathbf{r}_{L/2}-\mathbf{r}_{j})}\nonumber\\
&+\langle S_{L/2+1}^\alpha(0)S_{j}^\alpha(t) \rangle  e^{-i \mathbf{q}\cdot(\mathbf{r}_{L/2+1}-\mathbf{r}_{j})}\Big].\nonumber
\end{align}

We numerically determine the two-point spin correlation functions at different moments in time
\begin{align}
\label{eq:corr2}
\bra{0} S^\alpha_j (t) S^\alpha_{j'}\ket{0}=\bra{0} e^{i t H/\hbar} S^\alpha_j e^{-i t H/\hbar} S^\alpha_{j'}\ket{0},
\end{align}
where $\ket{0}$ the ground state of the spin model. We take the initial perturbation to be acting on the site $j'=L/2+1$, while the site $j$ covers the entire chain, i.e.~$j=1\dots L$. 
Afterwards, we perform the time-evolution of the state $\ket{\psi_{j'}}=S^\alpha_{j'}\ket{0}$ and compute the overlap between $S^\alpha_j\ket{\psi_{j'} (t)}$
and $e^{-i t E_0/\hbar}\ket{0}$, with $E_0$ the ground state energy.
The energy $E_0$ is taken from the ground state calculation of $\ket{0}$, as we do not need to perform its time evolution explicitly.
We perform the time-evolution only for positive times $t>0$, thus, we make us of the relation between the numerically computed correlations, Eq.~(\ref{eq:corr2}), and the ones necessary in Eq.~(\ref{eq:SF2}),
\begin{align}
\langle S_{j}^\alpha(t) S_{j'}^\alpha(0) \rangle=\langle S_{j}^\alpha(0) S_{j'}^\alpha(-t) \rangle=\langle S_{j}^\alpha(0) S_{j'}^\alpha(t) \rangle^*,
\end{align}
which assumes translational invariance, i.e.~the correlations only depend on the distance between the two sites, and that $S^\alpha$ is an Hermitian operator.
This leads to the following expression
\begin{align}
\label{eq:SF3}
\mathcal{S}^{\alpha\alpha}(\mathbf{q}, \omega)=&\sqrt{\frac{2}{L}}\Re\Bigg\{\sum_{j}\int_{0}^{\infty} dt\, e^{i \omega t} \\
\times&\Big[\langle S_{L+1-j}^\alpha(t) S_{L/2+1}^\alpha(0)  \rangle  e^{-i \mathbf{q}\cdot(\mathbf{r}_{L/2}-\mathbf{r}_{j})}\nonumber\\
&+\langle S_{j}^\alpha(t) S_{L/2+1}^\alpha(0) \rangle  e^{-i \mathbf{q}\cdot(\mathbf{r}_{L/2+1}-\mathbf{r}_{j})}\Big]\Bigg\}.\nonumber
\end{align}

The numerical Fourier transform to the frequency-momentum space is performed with the discrete frequencies $\omega=\frac{2\pi s}{N_t \delta t}$, where $N_t$ the number of time measurements up to the final evolution time and $\delta t$ the time interval between them, $s=0\dots N_t-1$. 
We apply a Gaussian filter in order to minimize the effects of the open boundary conditions before the numerical Fourier transform \cite{WhiteFeiguin2004, BouillotGiamarchi2011, HalatiBernier2023}
\begin{align}
\label{eq:corr_filter}
f(j)= e^{-4\left( 1-\frac{2j}{L-1} \right)^ 2}.
\end{align}
The Gaussian filter minimizes the numerical artifacts stemming from  the use of open boundary conditions. However, using the procedure will
reduces the momentum resolution. We chose the width of the Gaussian filter in order to balance between the two effects.
Putting all these together, the form of the dynamical structure factor we employ in the calculations is
\begin{align}
\label{eq:SF3}
\mathcal{S}^{\alpha\alpha}(\mathbf{q}, \omega)=&\sqrt{\frac{2}{N_t L}}\Re\Bigg\{\sum_{n=0}^{N_t-1} \sum_{j=1}^L e^{i \omega n\delta t}  \\
\times\Big[f(L+1-j)&\langle S_{L+1-j}^\alpha(t) S_{L/2+1}^\alpha(0)  \rangle  e^{-i \mathbf{q}\cdot(\mathbf{r}_{L/2}-\mathbf{r}_{j})}\nonumber\\
+f(j)&\langle S_{j}^\alpha(t) S_{L/2+1}^\alpha(0) \rangle  e^{-i \mathbf{q}\cdot(\mathbf{r}_{L/2+1}-\mathbf{r}_{j})}\Big]\Bigg\}.\nonumber
\end{align}

For a faithful comparison with the experimental results we need to use the momentum $\mathbf{q}$ of the neutrons and the positions of the ions $\mathbf{r}_{j}$ in the material, as we detail in the following.

\begin{figure}[!hbtp]
 \centering
 \includegraphics[width=0.45\textwidth]{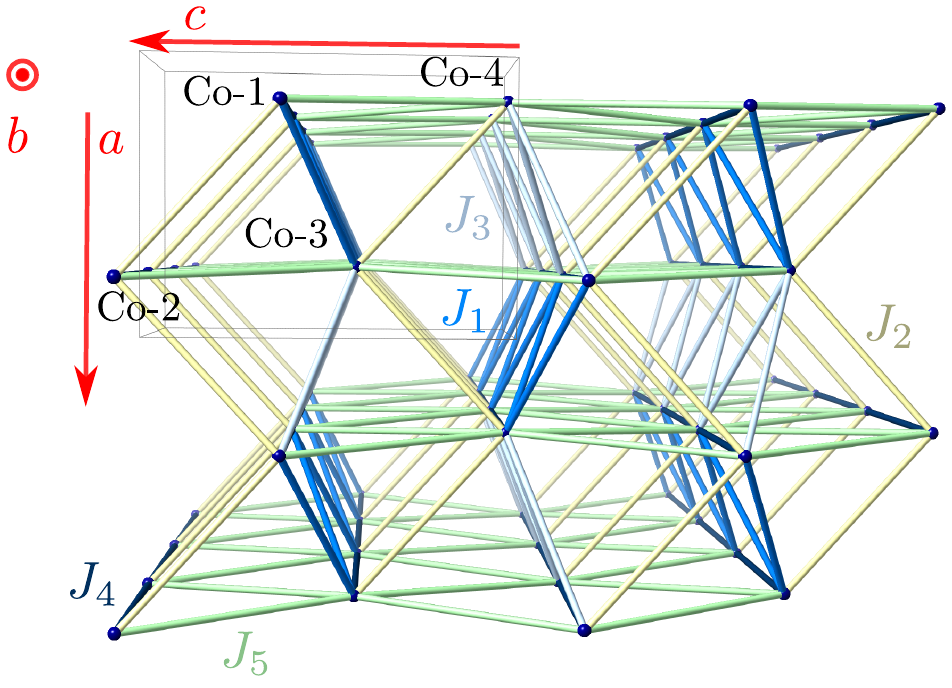}
 \caption{\label{fig:material}
Sketch of the five nearest-neighbor magnetic interactions in \CsCoBr. The nodes of the lattice are Co\textsuperscript{2+} ions, labeled within a unit cell by Co-1, Co-2, Co-3 and Co-4. Figure adapted from Ref.~\cite{FacherisZheludev2024}, where $J_4\equiv J$ in our notation of the exchange interactions.}
 \end{figure}

As seen in the sketch shown in Fig.~\ref{fig:material} the material consists in two different ladders, the first one labeled by the ions Co-1 and Co-3, and the second by Co-2 and Co-4.
In our theoretical analysis we consider only a mean-field coupling between the two ladders, thus, we can add independently the contributions to the dynamical structure factor stemming from the two ladders
\begin{align}
\label{eq:SF_MF}
\mathcal{S}^{\alpha\alpha}(\mathbf{q}, \omega)=\mathcal{S}_1^{\alpha\alpha}(\mathbf{q}, \omega)+\mathcal{S}_2^{\alpha\alpha}(\mathbf{q}, \omega),
\end{align}
where the subscript labels the two ladders.
The positions of the ions unit cell given as fractions of the lattice parameter are \cite{FacherisZheludev2024}
\begin{align}
\label{eq:positions}
\text{Co-1:}& (0.262,0.75,0.58)\\
\text{Co-2:}& (0.762,0.75,0.92)\nonumber\\
\text{Co-3:}& (0.738,0.25,0.42)\nonumber\\
\text{Co-4:}& (0.238,0.25,0.08).\nonumber
\end{align}
However, the ions that determine the ladder structures we are interested in are
\begin{align}
\label{eq:positions2}
\text{Co-1:}& (0.262,0.75,0.58)\\
\text{Co-2:}& (0.762,0.75,-0.08)\nonumber\\
\text{Co-3:}& (0.738,0.25,0.42)\nonumber\\
\text{Co-4:}& (1.238,0.25,0.08).\nonumber
\end{align}
Thus, in the theoretical results we can parameterize the positions of the ions as in the following, for the first ladder
\begin{align}
\label{eq:positions_L1}
&\mathbf{r}_j= \\
&\left(0.5+0.238 (-1)^j,0.25+0.5(j-L/2),0.5-0.08(-1)^j\right), \nonumber
\end{align}
and for the second ladder
\begin{align}
\label{eq:positions_L2}
&\mathbf{r}_j= \\
&\left(1+0.238 (-1)^j,0.25+0.5(j-L/2),0.08(-1)^j\right).\nonumber
\end{align}
These are in the units of lattice spacing $a=10.137~\text{\normalfont\AA}$, $b=7.593~\text{\normalfont\AA}$ and $c=13.281~\text{\normalfont\AA}$ \cite{FacherisZheludev2024}.

The momentum is parametrized as $\mathbf{q}=\left(\frac{2\pi h}{a},\frac{2\pi k}{b}, \frac{2\pi l}{c}\right)$. Usually, $h$ and $l$ are fixed and in the theoretical plots and we consider the momentum values spanned by $k=\frac{m}{L}$, with $m=0\dots L-1$.
This leads to the following scalar product between the momentum and position vectors
\begin{align}
\mathbf{q}\cdot (\mathbf{r}_j-\mathbf{r}_{j'})= 2\pi [ h(x_j-x_{j'}) +
k(y_j-y_{j'}) + l(z_j-z_{j'}) ].
\end{align}

The experimental results with polarized neutrons are taken at a fixed momentum, thus from the intensity given in Eq.~(\ref{eq:intensity}) we do not need the momentum dependent prefactors and we compare the signal obtain along the $\mathbf{b}$ direction with the theoretical result for $\mathcal{S}^{yy}$ and for the signal along the $\mathbf{a}$ direction the theoretical result for $\mathcal{S}^{xx}+\mathcal{S}^{zz}$ (as the angle between the $\mathbf{a}$ direction and $x$ spin direction is approximately $\pi/4$ in the $xz$-plane).

However, for the comparison with the unpolarized neutron scattering experimental measurements we need to compute the polarization factor and the form factor.
The crystallographic axes and the spin directions do not coincide, while $y|| \mathbf{b}$, $\mathbf{a}$ and $\mathbf{c}$ are rotated with respect to $x$ and $y$, in a different way for the two ladders forming the structure of the material.
For the first ladder, spanned by Co-1 and Co-3, we need to perform a rotation of $\pi/4$ around $\mathbf{b}$, and for the second ladder, spanned by Co-2 and Co-4,  we need a rotation of $-\pi/4$ around $\mathbf{b}$.
The rotation matrix around $\mathbf{b}$ is given by
$R_\mathbf{b}(\theta)=\begin{pmatrix}
  \cos(\theta) & 0 & -\sin(\theta)\\ 
  0 & 1 & 0 \\
  \sin(\theta) & 0 &\cos(\theta)
\end{pmatrix}$.
This implies that for an experimental momentum vector of 
\begin{align}
\label{eq:vec1}
\mathbf{q}=\left(\frac{2\pi h}{a},\frac{2\pi k}{b}, \frac{2\pi l}{c}\right),
\end{align}
in our theoretical calculations we need to consider for the first ladder
\begin{align}
\label{eq:vec2}
\mathbf{q}_{L1}=\left(\frac{2\pi}{\sqrt{2}}\left(\frac{h}{a}-\frac{l}{c}\right),\frac{2\pi k}{b}, \frac{2\pi}{\sqrt{2}}\left(\frac{h}{a}+\frac{l}{c}\right)\right),
\end{align}
and for the second ladder
\begin{align}
\label{eq:vec3}
\mathbf{q}_{L2}=\left(\frac{2\pi}{\sqrt{2}}\left(\frac{h}{a}+\frac{l}{c}\right),\frac{2\pi k}{b}, \frac{2\pi}{\sqrt{2}}\left(-\frac{h}{a}+\frac{l}{c}\right)\right),
\end{align}
as we computed the spin correlation functions along the spin directions and we have to rotate the momentum vector to the spin axes when computing the polarization factor and the magnetic form factor.
We note that the scalar products of the momentum and position vectors are left invariant under the rotations.

For our case the magnetic form factor is given by \cite{Brown2001}
\begin{align}
\label{eq:FF}
F(q)\propto \langle j_0(s)\rangle,
\end{align}
where $s=q/4\pi$ and
\begin{align}
\label{eq:FF2}
\langle j_0(s)\rangle=A_0e^{-a_0s^2}+B_0e^{-b_0s^2}+C_0e^{-c_0s^2}+D_0,
\end{align}
with the parameters $A_0=0.4332$, $a_0=14.3553$, $B_0=0.5857$, $b_0=4.6077$, $C_0=-0.0382$, $c_0=0.1338$, $D_0=0.0179$.

\setcounter{equation}{0}
\renewcommand{\theequation}{F.\arabic{equation}}
\setcounter{figure}{0}
\renewcommand{\thefigure}{F\arabic{figure}}

 \begin{figure}[!hbtp]
 \centering
 \includegraphics[width=0.45\textwidth]{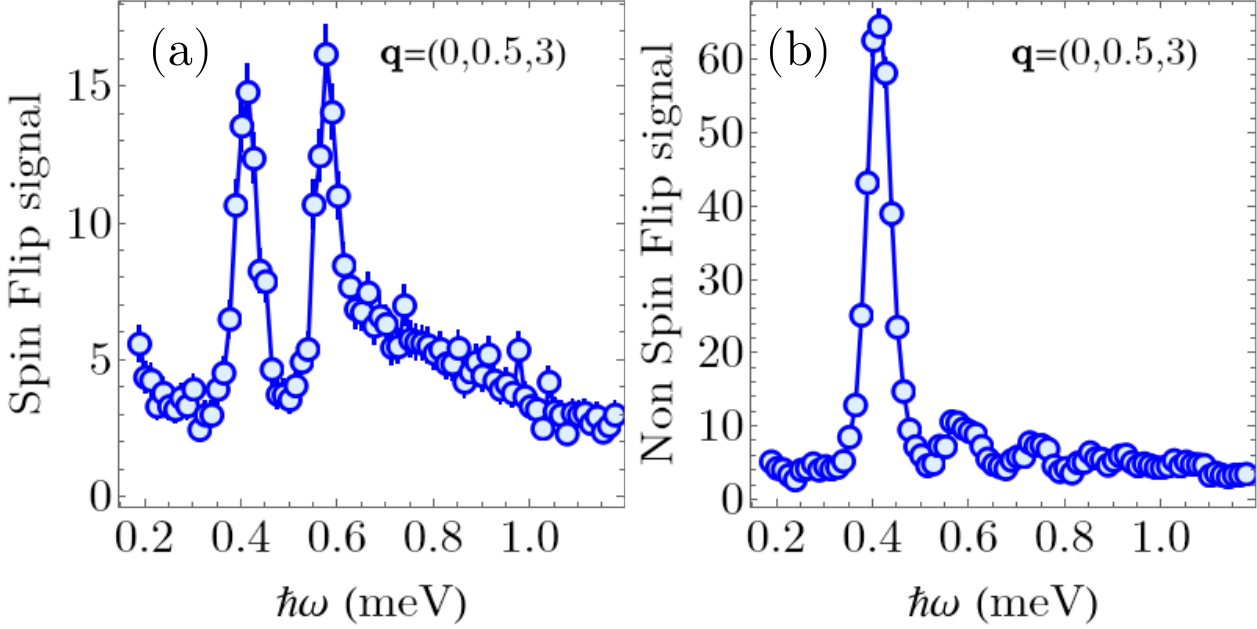}
 \caption{\label{fig:expdata}
Experimental intensity (arbitrary units) for inelastic polarized neutron scattering measurements for the compound Cs\textsubscript{2}CoBr\textsubscript{4} for the (a) Spin Flip signal and the (b) Non Spin Flip signal \cite{datapolarized}.}
 \end{figure}

\section{Experimental inelastic polarized neutron scattering data\label{sec:INS_data}}

In Fig.~\ref{fig:expdata} we show the raw experimental signal for the inelastic polarized neutron scattering measurements on the compound Cs\textsubscript{2}CoBr\textsubscript{4} \cite{datapolarized}, from which the data experimental data presented in Figs.~3(c)-(d) in the main text was obtained. The details of the experiment are presented in the main text.

\end{document}